\documentstyle[12pt,epsfig]{article}
\newcommand{\be}{\begin{equation}}
\newcommand{\ee}{\end{equation}}
\newcommand{\beq}{\begin{equation}}
\newcommand{\eeq}{\end{equation}}
\newcommand{\ba}{\begin{eqnarray}}
\newcommand{\ea}{\end{eqnarray}}
\newcommand{\bea}{\begin{eqnarray}}
\newcommand{\eea}{\end{eqnarray}}
\newcommand{\nn}{\nonumber}
\begin{document}
\baselineskip=15.5pt
\pagestyle{plain}
\setcounter{page}{1}


\def\del{{\partial}}
\def\vev#1{\left\langle #1 \right\rangle}
\def\cn{{\cal N}}
\def\co{{\cal O}}
\newfont{\Bbb}{msbm10 scaled 1200}     
\newcommand{\mathbb}[1]{\mbox{\Bbb #1}}
\def\IC{{\mathbb C}}
\def\IR{{\mathbb R}}
\def\IZ{{\mathbb Z}}
\def\RP{{\bf RP}}
\def\CP{{\bf CP}}
\def\Poincare{{Poincar\'e }}
\def\tr{{\rm tr}}
\def\tp{{\tilde \Phi}}

\def\TL{\hfil$\displaystyle{##}$}
\def\TR{$\displaystyle{{}##}$\hfil}
\def\TC{\hfil$\displaystyle{##}$\hfil}
\def\TT{\hbox{##}}
\def\HLINE{\noalign{\vskip1\jot}\hline\noalign{\vskip1\jot}} 
\def\seqalign#1#2{\vcenter{\openup1\jot
  \halign{\strut #1\cr #2 \cr}}}
\def\lbldef#1#2{\expandafter\gdef\csname #1\endcsname {#2}}
\def\eqn#1#2{\lbldef{#1}{(\ref{#1})}%
\begin{equation} #2 \label{#1} \end{equation}}
\def\eqalign#1{\vcenter{\openup1\jot
    \halign{\strut\span\TL & \span\TR\cr #1 \cr
   }}}
\def\eno#1{(\ref{#1})}
\def\href#1#2{#2}
\def\half{{1 \over 2}}

\def\ads{{\it AdS}}
\def\adsp{{\it AdS}$_{p+2}$}
\def\cft{{\it CFT}}

\newcommand{\ber}{\begin{eqnarray}}
\newcommand{\eer}{\end{eqnarray}}

\newcommand{\beqar}{\begin{eqnarray}}
\newcommand{\cN}{{\cal N}}
\newcommand{\cO}{{\cal O}}
\newcommand{\cA}{{\cal A}}
\newcommand{\cT}{{\cal T}}
\newcommand{\cF}{{\cal F}}
\newcommand{\cC}{{\cal C}}
\newcommand{\cR}{{\cal R}}
\newcommand{\cW}{{\cal W}}
\newcommand{\eeqar}{\end{eqnarray}}
\newcommand{\th}{\theta}
\newcommand{\lm}{\lambda}\newcommand{\Lm}{\Lambda}
\newcommand{\eps}{\epsilon}
\newcommand{\pa}{\paragraph}
\newcommand{\pt}{\partial}
\newcommand{\de}{\delta}
\newcommand{\De}{\Delta}
\newcommand{\lb}{\label}


\newcommand{\nonu}{\nonumber}
\newcommand{\oh}{\displaystyle{\frac{1}{2}}}
\newcommand{\dsl}
  {\kern.06em\hbox{\raise.15ex\hbox{$/$}\kern-.56em\hbox{$\partial$}}}
\newcommand{\id}{i\!\!\not\!\partial}
\newcommand{\as}{\not\!\! A}
\newcommand{\ps}{\not\! p}
\newcommand{\ks}{\not\! k}
\newcommand{\D}{{\cal{D}}}
\newcommand{\dv}{d^2x}
\newcommand{\Z}{{\cal Z}}
\newcommand{\N}{{\cal N}}
\newcommand{\Dsl}{\not\!\! D}
\newcommand{\Bsl}{\not\!\! B}
\newcommand{\Psl}{\not\!\! P}
\newcommand{\eeqarr}{\end{eqnarray}}
\newcommand{\ZZ}{{\rm \kern 0.275em Z \kern -0.92em Z}\;}

\begin{titlepage}

\leftline{\tt hep-th/0203124}

\vskip -.8cm

\rightline{\small{\tt CTP-MIT-3251}}
\rightline{\small{\tt DAMTP-2002-33}}

\begin{center}

\vskip 1.5 cm

{\LARGE RG flows from Spin(7), CY 4-fold and HK manifolds to AdS, 
Penrose limits and pp waves}
\vskip .3cm


\vskip 1.cm

{\large Umut G{\"u}rsoy$^1$$^{,a}$, Carlos N\'u\~nez$^2$, Martin Schvellinger$^1$$^{,b}$}

\vskip 0.6cm

$^1${\it Center for Theoretical Physics, \\
Laboratory for Nuclear Science and Department of Physics, \\
Massachusetts Institute of Technology, \\
Cambridge, Massachusetts 02139, USA} \\
$^a$ E-mail: {\tt umut@mit.edu} \\
$^b$ E-mail: {\tt martin@lns.mit.edu}

\vskip 0.5cm

$^2${\it Department of Applied Mathematics and Theoretical Physics, \\
Centre for Mathematical Sciences, University of Cambridge, \\
Wilberforce Road, Cambridge CB3 0WA, U.K.} \\
E-mail: {\tt C.Nunez@damtp.cam.ac.uk}


\vspace{1.7cm}

{\bf Abstract}
\end{center}

We obtain explicit realizations of holographic renormalization group (RG) flows 
from M-theory, from $E^{2,1} \times Spin(7)$ at UV to $AdS_4\times\tilde{S^7}$ (squashed $S^7$) at IR,  
from $E^{2,1} \times CY4$ at UV to $AdS_4 \times Q^{1,1,1}$ at IR, and 
from $E^{2,1} \times HK$ (hyperKahler) at UV to $AdS_4 \times N^{0,1,0}$ at IR.
The dual type IIA string theory configurations correspond to 
D2-D6 brane systems where D6-branes wrap supersymmetric four-cycles. 
We also study the Penrose limits and obtain the pp-wave backgrounds for the above configurations.
Besides, we study some examples of non-supersymmetric and
supersymmetric flows in five-dimensional gauge theories.


\noindent

\end{titlepage}

\newpage


\vfill

\section{Introduction}

Since M-theory compactifications on manifolds of special holonomy
preserve a fraction of the original supercharges in flat eleven-dimensional
spacetime, it has become a fruitful arena to explore the dynamical aspects of
minimally supersymmetric gauge theories.
Indeed, study of special holonomy manifolds developing an isolated classical
singularity has recently shed light on several important questions regarding 
restoration of global symmetries, phase transitions between different classical 
spacetimes \cite{Atiyah:2000zz,Atiyah:2001qf,
Gukov:2001hf,Acharya:2001hq,Cvetic:2001zb,
Acharya:2001dz,Gomis:2001vk,Cvetic:2001pg,pope,
Edelstein:2001pu,Brandhuber:2001yi,
Cvetic:2001cm,Gomis:2001vg,Cvetic:2000db,
Witten:2001uq,Cvetic:2001zx,Acharya:2001gy,
Blumenhagen:2001jb,Gomis:2001xw,Curio:2001dz,
Cvetic:2001kk,Cvetic:2001ih,Brandhuber:2001kq,
Sfetsos:2001ku,Cvetic:2001kp,Hernandez:2002fb,
Gukov:2002er,Gukov:2002es}, relations between anomalies
in M-theory, string theory and gauge theories \cite{Gukov:2001hf}, 
among other relevant aspects (see for instance \cite{Acharya:2001dz} to 
\cite{Hernandez:2002fb}). The dynamics of ${\cal {N}}=1$ SYM theory in four and 
three dimensions has been exhaustively investigated by Atiyah and Witten
\cite{Atiyah:2001qf} and Gukov and Sparks \cite{Gukov:2001hf}, respectively.

Particularly, it is possible to study certain properties of these backgrounds 
through their dual D6 brane configurations in type IIA string theory.
Any configuration of type IIA string theory with no bosonic content
except than the metric, Ramond-Ramond one-form and dilaton
lifts to an eleven-dimensional supergravity configuration without flux. 
This is a pure gravitational configuration. For instance, let us consider a
collection of $N_6$ parallel D6-branes in type IIA string theory \cite{Horowitz:cd}. 
In eleven dimensions the metric is described by the product of a seven-dimensional Minkowski 
spacetime and an Euclidean multi-centered Taub-NUT space \cite{Eguchi:1978xp}. 
Moreover, a configuration of D6-branes in flat space can be represented in M-theory by a
four-dimensional manifold with $SU(2)$ holonomy \cite{Gomis:2001vk}.
Furthermore, one can consider D6-branes wrapping supersymmetric cycles in spaces 
with special holonomy and, as described in \cite{Gomis:2001vk}, there are two different possibilities
that can be exemplified as follows. 
One can have D6-branes wrapping a supersymmetric four-cycle, $S^4$, in a $G_2$ holonomy manifold.
Thus, D6-branes completely fill the space transverse to type IIA string theory compactification
manifold, and therefore the field theory is on the transverse Minkowski three-dimensional 
spacetime, while the local M-theory description involves a Spin(7)
holonomy manifold.
As another example, one can consider D6-branes wrapping a different four-cycle,
$S^2 \times S^2$, in a CY3 fold. Then, the three-dimensional field theory is codimension
one in the transverse Minkowski space to the type IIA compactification
manifold. In this case, the local M-theory description is given by
a CY4 fold. The corresponding pure eleven-dimensional geometric configurations 
were obtained long time ago in \cite{spin7} and \cite{calabiyau}.
Recently, a supergravity solution was obtained when D6-branes 
are wrapped on $S^4$ in seven-dimensional manifolds of $G_2$ holonomy \cite{Hernandez:2001bh}. 
This solution preserves two supercharges and thus it represents a supergravity dual of a three-dimensional 
${\cal {N}}=1$ SYM theory. Lifted to eleven dimensions this solution describes M-theory on the background 
of a Spin(7) holonomy manifold. A detailed analysis of the dual field theory has been done in
\cite{Gukov:2001hf}. In addition, 
supergravity duals of D6-branes wrapping Kahler four-cycles inside a CY3 fold have been obtained
in \cite{Gomis:2001vg}. In this case the 
purely gravitational M-theory description corresponds to a CY4 fold. 

A natural step forward in these investigations
is to explore the role of the background four-form field strength
in compactifications of M-theory on manifolds of special holonomy. 

Existence of $F_4$ field strength will deform the geometry into a different
background. In this paper we will study the situation when $F_4$ flux
is taken on the three-dimensional Minkowski space-time plus the radial coordinate.
Some questions that can be addressed are the geometry induced by this
$F_4$ flux, the dynamical mechanism to turn on $F_4$ field strength 
and the relations among the topological cycles in M-theory,
type IIA string theory and field theory in such backgrounds.
The natural frame to ask these 
questions is eleven-dimensional supergravity\footnote{Also, considering the 
standard issues in the duality between type IIA string theory and
eleven-dimensional supergravity, the correspondence between certain degrees of freedom in type IIA string
theory and M-theory gives evidence to suspect that this duality goes beyond supergravity approximation
\cite{Gubser:2002mz}.}. Since the corresponding gauge theory on D6-branes 
is a seven-dimensional one, 
it is actually more natural to find supergravity solutions in a simpler
eight-dimensional gauged supergravity \cite{Salam:1984ft}.
Therefore, we will find the dynamical behavior of $F_4$ (in the ``flat directions'')
by solving eight-dimensional supersymmetric configurations. Then, we
will perform the uplifting to eleven dimensions and study
holographic RG flows in three situations.  
One from $E^{2,1} \times Spin(7)$ at UV to
$AdS_4\times\tilde{S^7}$ (squashed $S^7$) at IR, 
which corresponds to the first case in the classification of \cite{Gomis:2001vk}.
A second case will correspond to a flow from 
$E^{2,1} \times CY4$ fold at UV towards $AdS_4 \times Q^{1,1,1}$ in the IR limit.
Finally, we will consider the case of 
$E^{2,1} \times HK$ at UV towards $AdS_4 \times N^{0,1,0}$ in the IR limit.
In the IR limit, they represent duals of ${\cal {N}}=1, \, 2$ and $3$ super Yang Mills
theories in three dimensions, respectively. 
The system under study consists of 
localized D2-branes inside D6-branes.
We will see that, as the theory flows to the IR limit, $F_4$
through the ``flat directions'' dynamically increases. However,
the number of localized D2-branes inside D6-branes remains constant,
hence leading to a D2-D6 brane system.
We leave the issue of exploring dynamics of the four-form field
strength which lives on the four-cycle coordinates for a future investigation. An important study 
regarding this last $F_4$ configuration has been addressed in
\cite{Acharya:2002vs}, although without discussing the corresponding supergravity duals.

Very recently, Berenstein, Maldacena and Nastase have proposed a compelling idea
explaining how the string spectrum in flat space and pp-waves arise from the large $N$ limit
of $U(N)$ ${\cal {N}}=4$ super Yang Mills theory in four dimensions at fixed $g_{YM}$
\cite{Berenstein:2002jq}.
This idea has been applied to some different backgrounds 
\cite{Metsaev:2002re,Blau:2002rg,Berenstein:2002ke,
Itzhaki:2002kh,Gomis:2002km,Russo:2002rq,
Zayas:2002rx,Hatsuda:2002xp,Alishahiha:2002ev,Billo':2002ff,
Kim:2002fp,Cvetic:2002hi,Takayanagi:2002hv,Floratos:2002uh}.
For all of the IR backgrounds mentioned above we will study 
the corresponding Penrose limit and
obtain their pp-wave background. Interestingly, in each case we find
the enhancement of supersymmetry 
from ${\cal {N}}=1, \, 2$ and $3$ to ${\cal {N}}=8$ in the dual three-dimensional SYM theory
in the Penrose limit. 
Our examples support a similar enhancement phenomenon already found
for ${\cal {N}}=1$ to ${\cal {N}}=4$ super Yang Mills theory in four dimensions 
\cite{Itzhaki:2002kh,Gomis:2002km}.  

The paper is organized as follows. In the next section we will describe the general idea and
motivations. In section 3 we describe some generalities of the D2-D6 brane
system in the flat case. In section 4 we obtain an RG flow from Spin(7) holonomy manifolds
at UV to AdS spaces at IR and, also discuss the field theory duals. Then, in section 
5 we will consider flows from CY4 folds to AdS spaces and a case
preserving ${\cal {N}}=3$ supersymmetries in three dimensions. 
Section 6 is devoted to an analysis of the Penrose limits in the IR
region of the supergravity solutions mentioned above.   
In section 7 we will study some examples of 
non-supersymmetric and supersymmetric flows in five-dimensional gauge theory.
These flows will also be of interest since their uplifting to massive type IIA
is known \cite{cveticmassive}. Appendix A introduces eight-dimensional gauged
supergravity and discusses its relevant aspects related to our present interest. 
In Appendix B we present more general super-kink solutions of BPS
equations which include above as special cases and consider their
dual RG flow interpretation. In Appendix C we show some numerical solutions.
Finally, in Appendix D we introduce some notation for the squashed seven-sphere.

\section{General idea}

As mentioned in the introduction, we will find supergravity solutions
describing the RG flow from a special holonomy manifold (Spin(7) or CY4
folds) to manifolds of the form $AdS_4 \times \tilde{M_7}$. We can
understand these flows by realizing the fact that, since three-dimensional
gauge theories have a dimensionful coupling constant, they flow to
interacting IR fixed points \cite{Pelc:1999ms,Intriligator:1996ex,deBoer:1996mp}.  
These flows are interesting by their own,
since they realize new examples of AdS/CFT correspondence and some
generalizations of it.

The way in which we will find our solutions is the following: we will
start from the eight-dimensional $SU(2)$ gauged supergravity \cite{Salam:1984ft},
that was proven to descend from eleven-dimensional supergravity as a reduction
on $S^3$ (where only one of the two $SU(2)$s is being gauged). We will find the solutions in
this lower dimensional supergravity, and then lift them to eleven 
dimensions.
  
The advantage of doing the computations in this way is that,
working with a delocalized D2-D6 system, in principle, one has
to deal with a seven-dimensional gauge theory, hence one is naturally
led to consider an eight-dimensional gravity theory. Indeed, we will see that
after lifting, our solutions represent either D6 branes or a system of D2-D6
branes. Then, we will wrap D6 branes on some supersymmetric
cycle, leading to a localized D2-D6 system. 
As it is well known, when a brane wraps a supersymmetric cycle, there is a
way to preserve some amount of supersymmetry through the so-called twisting mechanism
\cite{Bershadsky:1995qy}. Realization of this mechanism in supergravity 
is basically the equality (here we suppress gamma matrices) of
the spin connection of the manifold and the gauge field of the gauged
supergravity under study, {\it i.e.} $\omega_\mu= A_\mu$, such that this combination
is canceled in the covariant derivative, and thus allowing one to define
a covariantly constant Killing spinor everywhere on the brane.
Many interesting realizations of this mechanism have been previously worked out
(see \cite{maldacena-nunez} to \cite{last-twisting}).

We will construct solutions where D6-branes are wrapping a four-cycle
($S^4$) inside a $G_2$ holonomy manifold, and a second set 
of solutions where D6-branes wrap a four-cycle ($S^2 \times S^2$) inside a
CY3 fold. Also, we consider an example preserving ${\cal {N}}=3$ supersymmetries in three
dimensions. These examples realize M-theory 
configurations preserving ${\cal {N}}=1$,
${\cal {N}}=2$ and ${\cal {N}}=3$  supersymmetries in three dimensions,
{\it i.e.} two, four and six supercharges 
respectively.

\section{The system under study}

As mentioned above, we will firstly study a delocalized D2-D6 brane system.
In order to see this explicitly from a metric description, let us
construct solutions in eight-dimensional supergravity where the field
content will be a dilaton $\phi(r)$, a four-form field $G_4$ and a metric
of the form, 
\beq
ds_8^2=e^{2f} \, dx_{1,2}^2 + dr^2  + e^{2h} \, d\vec{y}_4^{\, 2} \,\,\, ,
\label{metrica1}
\eeq
\beq
G_{x_1 x_2 x_3 r}= \Lambda \, e^{-4 h - 2 \phi} \,\,\, ,
\label{F41}
\eeq
where $dx_{1,2}^2$ is the flat Minkowski metric in $3$ dimensions.
In Eq.(\ref{F41}) we have written the four-form field in flat indices. In that
follows we will assume the scalar functions $f$, $h$, $\phi$ (and also $\lambda$)
to be only $r$-dependent. 

Plugging this configuration into the supersymmetric variations of the fermion fields
and requiring these variations to vanish, one can obtain a system of BPS equations
(where prime denotes derivative with respect to $r$)
\ba
f'    &=&  - \frac{1}{8} \, e^{-\phi} - 
\frac{\Lambda}{2} \, e^{-4 h -\phi} \,\,\, ,
\label{fprime1} \\
\phi' &=& - \frac{3}{8} \, e^{-\phi} +
\frac{\Lambda}{2} \, e^{-4 h -\phi} \,\,\, ,
\label{phiprime1} \\
h'    &=& - \frac{1}{8} \, e^{-\phi} +
\frac{\Lambda}{2} \, e^{-4 h -\phi} \,\,\, .
\label{hprime1}
\ea
Following the prescription given in ref.\cite{Salam:1984ft}, one can easily see
that after lifting the solutions of the system above, they will correspond to M-theory 
configurations of the form
\beq
ds_{11}^2=e^{2f-2 \phi/3} \, dx_{1,2}^2 + e^{-2 \phi/3} \, dr^2
+ e^{2h-2 \phi/3} \, d\vec{y}_4^{\, 2} + 4 \, e^{4 \phi/3} \, d\Omega_3^2 \,\,\, ,
\label{metrica2}
\eeq
\beq
F_{x_1 x_2 x_3 r}= 2 \, \Lambda e^{-4 h - 2 \phi/3} \,\,\, ,
\label{F42}
\eeq
where again we have used flat indices for the four-form field strength.

Now, we want to interpret the equations above as describing a D2-D6 brane system.
Indeed, by setting $\Lambda$ equal to zero, the solution is given by the metric
corresponding to D6-branes in the near horizon region (lifted to M-theory)
\cite{Boonstra:1998mp}.
On the other hand, if we consider non-vanishing $\Lambda$, we can compute
a solution that shows the presence of D2-branes delocalized inside the 
D6-brane worldvolume. In this case the M-theory solution is
\beq
ds_{11}^2= \frac{\rho^2}{36} \, dx_{1,2}^2 +
\frac{4 \, \sqrt{\Lambda}}{9 \, \rho} \, dy_4^2 + 
9^3 \, d\rho^2 + 81 \, \rho^2 \, d\Omega_3^2 \,\,\, ,
\eeq
\beq
F_{x_1 x_2 x_3 \rho}=\frac{1}{18 \, \rho} \,\,\, .
\eeq
This solution is the near horizon limit of the one obtained in
\cite{Jarv:2000zv}.

\section{From Spin(7) holonomy manifolds to AdS spaces and FT duals}

\subsection{D6-branes wrapping cycles inside $G_2$ holonomy manifolds}

In this section we will consider D6-branes wrapping a four-sphere
in a $G_2$ holonomy manifold\footnote{We consider non-compact manifolds with
special holonomy. For the construction of compact manifolds with special 
holonomy see \cite{joyce}.}.
Furthermore, we will add D2-branes in the unwrapped
directions. After the twisting is performed
we obtain a $2+1$-dimensional gauge theory with 2 supercharges, {\it i.e.} 
$SU(N)$ ${\cal {N}}=1$ SYM theory in three dimensions. 

Our solution will describe a flow of this theory from a Spin(7) holonomy manifold 
to an AdS manifold, thus realizing the flow towards the IR fixed point that this 
kind of theories has, but in a gravitational set-up.
We found a second class of solutions which we included in Appendix B. 
However, these have singularities 
which makes investigation of the field theory duals
more difficult. 

As we have already mentioned, when D6-branes wrap a cycle, 
in order to preserve some amount of supersymmetries we have to do a twisting
procedure in the seven-dimensional gauge theory.  
In the present case we perform the twisting with
an $SU(2)$ gauge connection, and choosing the four-cycle to be a
four-sphere, we will have an $SU(2)$ instanton on $S^4$.

As it is well known, when we wrap D6-branes 
on a curved cycle there will be an induced D2-brane charge, that can be understood as coming
from the WZ coupling in the Born-Infeld action \cite{Green:1996dd}, of the form
\beq
\int_{E^{2,1} \times S^4} C_3 \wedge (F_2 \wedge F_2 + R_2 \wedge R_2) \,\,\, .
\eeq
The second term leads to an induced D2-brane charge similar to the one
that we propose in our configuration and it can be understood as an effective cosmological constant. 
When we turn on an $F_4$ flux in the four-cycle, the first term induces a Chern-Simons
term in the $2+1$ field theory. We postpone the study of 
this last type of interesting configurations for the future.

Let us consider a metric description for this field configuration.
In eight-dimensional supergravity we can use the following metric ansatz 
\beq
ds_8^2=e^{2f} \, dx_{1,2}^2 + dr^2 + e^{2h} \, d\Omega_4^2 \,\,\, ,
\eeq
while in flat indices the four-form field strength is defined as in Eq.(\ref{F41}).
Then, the BPS equations become
\ba
f' &=& \frac{1}{8} \, e^{-\phi} - e^{ \phi-2 h} + 
\frac{\Lambda}{2} \, e^{-4 h -\phi} \,\,\, , 
\label{fprime} \\
\phi' &=& \frac{3}{8} \, e^{-\phi} -3\,\, e^{ \phi-2 h} - 
\frac{\Lambda}{2} \, e^{-4 h -\phi} \,\,\, , 
\label{phiprime} \\
h' &=& \frac{1}{8} \, e^{-\phi} +2\,\, e^{ \phi-2 h} - 
\frac{\Lambda}{2} \, e^{-4 h -\phi} \,\,\, .
\label{hprime}
\ea
Since when $\Lambda=0$ the system above reduces to the one studied in
\cite{Hernandez:2001bh}, we will have the same solution, namely a cone over weak $G_2$. 
This solution is singular at IR and this singularity can be 
resolved by considering more elaborate solutions, in our case, this is
basically attained by
including an integration constant, such that the complete solution reads
\beq
ds^2= dx_{1,2}^2 + \frac{9}{20}\rho^2 d\Omega_4^2
+ \frac{9\rho^2}{100} \left(1 - \left(\frac{a}{\rho}\right)^{10/3}\right) \, (\omega-
A)^2 + \frac{d\rho^2}{\left(1 - \left(\frac{a}{\rho}\right)^{10/3}\right)} \,\,\, ,
\eeq
which is the metric of a Spin(7) holonomy manifold having the topology of an ${\bf R}^4$ bundle over $S^4$.

After the appropriate modding out by $Z_N$, this metric describes 
the M-theory version of the gravitational side of the 
geometrical transition between $N_6$ D6-branes wrapping a four-cycle inside a $G_2$ manifold
and a situation without branes but flux over a two-cycle.
Solutions similar to these ones, but with a stable $U(1)$ at long distances, 
were recently studied in \cite{Cvetic:2001pg,pope,Cvetic:2001cm}.

We can try to achieve a different type of resolution, by turning on another
degree of freedom in M-theory, namely the $F_4$ field, that is $\Lambda$ being non-zero 
in the system (\ref{fprime})-(\ref{hprime}). Obviously, this
will take us out from the ``pure metric'' configurations, but the type of resolution 
is well-behaved and it describes a phenomenon
that seems to be a common characteristic in three-dimensional gauge theories.
In the following, we are going to introduce different solutions of the above BPS equations, and discuss
them in terms of their ability in describing physically meaningful holographic RG flows.

\newpage

{\it A solution driving the flow from $E^{2,1} \times Spin(7)$ to $AdS_4\times\tilde{S^7}$}

Through the following change of variables 
\be
dr = e^{\phi} \, d\tau \,\,\, ,
\label{rt}
\ee
one can find a solution that reads 
\ba
h(\tau)&=& \frac{1}{4} \, \log \left(\frac{20 \, \Lambda}{9} + \frac{A}{9} \, e^{9\tau/10}\right) 
\, , \;\;\;\;\;\;\;\;\;\; \phi(\tau)= -1/2 \, \log(20) + h(\tau) \, , \nn \\ 
f(\tau) &=& \frac{3 \tau}{10} - \frac{1}{4} \, \log \left(20 \, \Lambda + A \, 
e^{9 \tau/10}\right) \,\,\, ,
\label{sol1}
\ea
where $A$ is an integration constant. The corresponding eleven-dimensional metric is
\ba
ds_{11}^2 & = & \frac{60^{1/3} \, e^{3 \tau/5}}{(A \, e^{9 \tau/10} + 20 \, \Lambda)^{2/3}}
dx_{1,2}^2 + \left( \frac{2\sqrt{5}}{3} \right)^{2/3}  \, 
(A \, e^{9 \tau/10} + 20 \, \Lambda)^{1/3} \, d\Omega_4^2 \nn \\ 
&& + (\frac{2}{15})^{2/3} \, (A \, e^{9 \tau/10} + 20 \, \Lambda)^{1/3} \, 
(\omega^i - A^i)^2 + \frac{1}{(2 \sqrt{15}) ^{4/3} } \, (A \,  e^{9 \tau/10} 
+ 20 \,  \Lambda)^{1/3} \, d\tau^2 \, , \nn \\
&&
\label{metriccompleta}
\ea
while in flat indices the four-form field strength is
\be
F_{x_1 x_2 x_3 \tau}= \frac{18 \,(2 \sqrt{15})^{2/3}  \, \Lambda}
{(A \, e^{9 \tau/10} + 20 \, \Lambda)^{7/6}} \,\,\, .
\ee
In order to obtain the M-theory configuration we have used the Salam-Sezgin's prescription 
to lift the metric and the four-form field strength to eleven
dimensions. It is interesting to note that
although we obtained the solution (\ref{metriccompleta})
from an eight-dimensional perspective, it can be recognized as the well-known 
M2-brane solution with fewer supersymmetries to eleven-dimensional supergravity whose
existence was first proved in \cite{Duff:1996} and, its explicit form was
obtained in \cite{Ferrara:1998}. Equivalence of solution (\ref{metriccompleta}) with the
M2-brane solution can be seen by changing the radial variable as
$r\,=\,e^{3\tau/20}$. This is further checked below by comparing the mass
spectrum obtained from eight-dimensional BPS equations in the IR limit with the
spectrum of $AdS_4\times\tilde{S^7}$ compactification of eleven-dimensional
supergravity \cite{Pope:1983}. 
 
Let us now try to understand the two regimes described by this metric, by analyzing the UV and IR
limits. For large values of $\tau$ we have
$\lim_{\tau \to  +\infty} h(\tau) = +\infty, $
leading to a large radius for the four-sphere. Besides, the 
Ricci scalar $R$ vanishes as $\tau \to 
+\infty$. Changing variables again\footnote{We use $\rho = \frac{20}{3} \, 
\frac{A^{1/6}}{2^{2/3} \, (15)^{1/3}} \, e^{3 \tau/20}$.}
in the limit of large $\tau$
the metric (\ref{metriccompleta}) becomes
the cone over the squashed (weak $G_2$) Einstein seven-sphere,
{\it i.e.} the large distance limit of the Spin(7) holonomy 
manifold given in \cite{Hernandez:2001bh,spin7,Gukov:2001hf}
\be
ds_{11}^2  =  dx_{1,2}^2 + d\rho^2 + \frac{9}{20} \, \rho^2 \, d\Omega_4^2 +
\frac{9}{100} \, \rho^2 \, (\omega^i - A^i)^2 \,\,\, ,
\label{spin7cone}
\ee
while, of course,
$F_{x_1 x_2 x_3 r}=0$. In figure 1 we show the behavior of 
$F_{x_1 x_2 x_3 r}$, $2/3 F_{x_1 x_2 x_3 r}^2$ and the Ricci scalar $R$, 
as a function of $\tau$. We can see how the solid lines are exactly the same curve up to
a minus sign. This graphically reflects the validity of the eleven-dimensional equations of motion.

\vspace{0.5cm}
\begin{center}
\epsfig{file=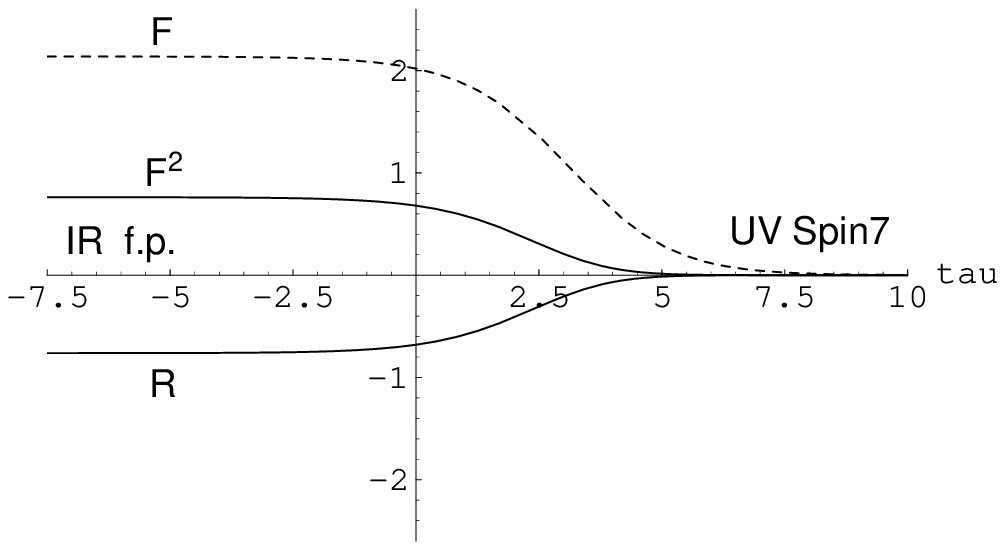, width=9cm}
\end{center}
\vspace{0.5cm}
\baselineskip=13pt
\centerline{\small{Figure 1: $F_{x_1 x_2 x_3 r}$, $2/3 F_{x_1 x_2 x_3 r}^2$ 
(labeled as $F^2$) and the Ricci scalar $R$,}}
\centerline{\small{as a function of $\tau$.}}

\vspace{0.5cm}

\baselineskip=15.5pt

From this figure, we can also see how both the four-form field strength and the Ricci scalar
go to zero at the UV limit. From the vanishing of the
Ricci scalar it shows how the eleven-dimensional manifold becomes flat at UV, while in 
the IR limit it is negative, as one expects from AdS-like spacetimes.

We can understand  some aspects of the field theory at the UV regime described by the
metric (\ref{spin7cone}). A very beautiful paper analyzing these kind of aspects
is \cite{Gukov:2001hf}. We can understand topological objects in the
effective $2+1$-dimensional field theory
as follows. From an M-theory
perspective they must correspond to M2 or M5-branes wrapping non-contractible cycles, 
that ``intersect'' the $2+1$-dimensional worldvolume.
Existence of suitable non-trivial cycles will signal the possibility of 
having a given topological defect. For example, 
a domain wall will be represented by an M5-brane wrapping 
a four-cycle, so the existence of domain walls
will be determined by non-triviality of $H_4(X_8,Z)$ (where $X_8$ 
is the eight-dimensional manifold ``external'' to 2+1 flat directions).
Monopoles and instantons will be associated with $H_5(X_8,Z)$ and $H_3(X_8,Z)$
since they will correspond to M5-branes
wrapping five-cycles and membranes wrapping three-cycles.
Indeed, there is a correspondence between M-theory and type IIA string theory
degrees of freedom. For instance, if one considers the multi-centered 
Taub-NUT metric times a seven-dimensional Minkowski spacetime in eleven dimensions,
in type IIA string theory one can think of that as $n+1$ parallel 
D6-branes placed at each center $r_i$ of the Taub-NUT four-dimensional metric.
In eleven dimensions the $A_n$ singularity can be resolved by $n$ homologically non-trivial 
cycles at $r_i$. Therefore, there are $n$ normalizable cohomological two-forms, $\omega^i$. 
Also there is an additional normalizable two-form $\omega^0$ with no topological meaning.
Then, the expansion of the Ramond-Ramond three-form of type IIA string theory can be done as
\be
C_{(3)} = \sum_{i=0}^n \, \omega^i \wedge A_i \,\,\, .
\ee
It involves an $U(1)$ seven-dimensional gauge field localized at the center of the Taub-NUT space,
and it corresponds to the $U(1)$ gauge field on each D6-brane. In addition, there are 
$n(n+1)/2$ holomorphic embeddings of two-cycles in the mentioned Taub-NUT metric, and an M2-brane
wrapped on any of these is a BPS particle, while in type IIA string theory 
it becomes a string stretched between two D6-branes \cite{Gubser:2002mz}.

As explained in \cite{Gukov:2001hf}, metrics like the one in
Eq.(\ref{metriccompleta}) are very good classical backgrounds, but they fail to give a good description
of the quantum theory. Indeed, they suffer membrane anomalies. The absence of $F_4$ 
flux in the curved part ($S^4$, $S^2 \times S^2$ and a $CP^2$ manifold, in this paper) 
turns out to be the source of global membrane
anomalies \cite{Freed:1999vc,Witten:1996md}. Our metrics do not cure this problem,
however they are a step towards its resolution by dynamically turning on the $F_4$ field. 
In this case this flux gives a number of localized 
D2-branes inside D6-branes in type IIA string theory.
In eleven-dimensional supergravity this number of M2-branes (following the
notation of \cite{Gukov:2001hf} we will denote it by $N_{M2}$)
will have a very interesting effect on the system. These D2-branes are 
instantons from the viewpoint of the ${\cal {N}}=4$, $d=4$ twisted topological 
super Yang Mills theory living
on the curved part of the D6-branes, hence they must encode the information
of the moduli space of $N_{M2}$ instantons. 

When the number of D2/M2-branes is zero (the case when $\Lambda=0$ in our
BPS system) there are no dynamical  scalars
in the worldvolume theory. This can be seen in two ways. 
First, one observes that D6-branes wrapping a four-cycle in a $G_2$ holonomy
manifold do not leave us with flat transverse
directions where the brane would in principle fluctuate. Also, no
hypermultiplets are present since there are no massless 
modes that could be excited on the four-sphere. 
Secondly, following \cite{Atiyah:2001qf}
one can compute the fluctuations of the metric and see that they are not square
integrable in the eight-dimensional manifold, rendering the
fluctuations non-dynamical which should be interpreted as a coupling constant. 
As we will see in the next section, this situation changes when we consider
the ${\cal {N}}=2$ version of this set up. In that case one real
scalar field will be dynamical and together with
the vector field, it will fill the ${\cal {N}}=2$ supermultiplet.

Nevertheless, as mentioned above, in our system where the $C_3$ field is excited
additional dynamical scalars will appear. These scalars will encode the
information on the instantons in the topological field theory on $S^4$. Indeed, 
these scalars do not have a purely geometrical origin. If this were the case, 
they would not have dynamics and should just represent a coupling constant.

Now, we should clarify a point. In the UV limit, we do not expect the theory to be strictly
a three-dimensional one, since likely at very high energies there will 
be massive modes excited on the cycle $S^4$, however, one can think about 
that as doing a fine-tuning
of the constants of the problem, that is the strength ($\Lambda$) of the $F_4$ field 
and the integration constant $A$ such that, even when we are at high energies, the theory is nearly 
three-dimensional. In any case the existence of the topological objects described above 
and other features are independent of the scale of energy at which we are observing the theory.

Our complete metric Eq.(\ref{metriccompleta}) describes the flow between a theory with 
the characteristics mentioned
above and a supersymmetric conformal field theory, described by
the fixed point solution we study bellow. The gravitational background is that of 
M2-branes on the tip of an Spin(7) cone. In order
to see this,  we will consider the situation 
in which the radial variable $\tau$ takes very large negative values ($\tau\to -\infty$)
and we can see  in this limit how the solution reaches the fixed point, {\it i.e.} 
$AdS_4 \times \tilde{S^7}$, 
\be
ds_{11}^2  =  \left( \frac{3}{20 \, \Lambda^2} \right)^{1/3} \, e^{3 \tau/5} \, dx_{1,2}^2 +  
\left( \frac{\Lambda}{180} \right)^{1/3} \,  d\tau^2
+ \left( \frac{20}{3} \right)^{2/3} \, \Lambda^{1/3} \, d\Omega_4^2  
+  \left( \frac{16 \Lambda}{45} \right)^{1/3} \, 
(\omega^i - A^i)^2 \,\,\, .
\label{metricfp}
\ee
As usual $\omega^i$ are the left-invariant one-forms defined on the 
$SU(2)$ group manifold, while $A^i$
corresponds to an $SU(2)$-instanton with only non-vanishing components on the four-sphere.
Both the Ricci scalar and the four-form field strength become constant
\be
F_{x_1 x_2 x_3 \tau}= \frac{9 \, \cdot \, 3^{1/3}}
{2^{2/3} \, 5^{5/6} \, \Lambda^{1/6}} \,\,\, .
\ee

It is staightforward to work out the scale dimensions of the operators
in the corresponding CFT at IR. This is most easily done by
linearizing Eqs.(\ref{hprime}) and (\ref{phiprime}) near the fixed point
\be
h= \frac{1}{4} \, \log \left(\frac{20 \, \Lambda}{9}\right) 
\, , \;\;\;\;\;\;\;\; \phi= -1/2 \, \log(20) + h \, . 
\label{fp1}
\ee
The eigenvectors of the mass matrix turn out to be, $\varphi_1\,\equiv\,\phi\,-\,h$ and 
$\varphi_2\,\equiv\,\phi\,+\,6\,h$ whose linearized solutions read,
\be
\varphi_1(r)\,\,\approx\,\, A_1\,e^{-\frac{4}{3}\,r}\, , \;\;\;\;\;\;\;\; 
\varphi_2(r)\,\,\approx\,\, A_2\,e^{3\,r}\,  
\label{fpsol1}
\ee
which shows that scale dimensions of the correspoding operators in dual
CFT read,
\be
\Delta_{\varphi_1}\,\,=\,\,6\, ,
\;\;\;\;\;\;\;\;\Delta_{\varphi_2}\,\,=\,\,\frac{4}{3}\,\,
{\mathrm{or}}\,\,\frac{5}{3}\,.
\ee
The ambiguity in the second dimension is due to two possible
unitary deformations of the CFT with a source or VEV \cite{Klebanov:1999}. The
corresponding masses are $M^2_1\,=\,-\frac{80}{9}\,m^2$,
$M^2_2\,=\,+72\,m^2$ in mass unit $m^2$ set by the radius of
$AdS_4$. One can easily see from the uplifting formulae in Appendix A
that, $\varphi_1(r)$ and  $\varphi_2(r)$ correspond to ``squashing''
and ``overall size'' deformations of $\tilde{S^7}$, respectively. Therefore, we see
that ``squashing'' is a \emph{relevant} whereas ``overall
size'' is an \emph{irrelevant} deformation in the dual CFT. In 
\cite{AhnRey:1999} the squashing operator, $\varphi_1$
was identified with the singlet of the isometry group $SO(5)\times SO(3)$ of
$\tilde{S^7}$ under the decomposition of $(0,\,2,\,0,\,0)$ of $SO(8)$.
Obviously the overall size operator, $\varphi_2$
should be  identified with the singlet of $SO(5)\times SO(3)$
inherited from $(0,\,0,\,0,\,0)$ of $SO(8)$. Looking at
the standard tables for mass operators, {\it e.g.} in
\cite{DuffPopeNilsson:1986}, one indeed sees that it is the singlet in
the KK-tower $0^{+(3)}$ and has $M^2_2\,=\,+72\,m^2$ and energy
$\Delta=6$. Note that in our solution (\ref{sol1}) we turned off the
squashing deformation, {\it i.e.} we set $h\,-\,\phi\,=\,const$. This is a
supersymmetric hence stable truncation. Since the only relevant
deformation in the dual theory is turned off and we are left with an
irrelevant overall size deformation, CFT appears in IR. On the other
hand, in the more general solution of Appendix B where we turn on
$\varphi_1$ along with $\varphi_2$, dual CFT appears in the UV limit, as
expected.    
 
It is interesting to plot the solution in terms of the variables
introduced in \cite{Acharya:2000mu},
$H=s^2 e^{-2 \phi}$ and $s=e^{2 h}$. 
These variables are choosen in order to study the orbits of the ODEs 
\be
\frac{d H}{d s} = \left(\frac{ - H \,\, s^2 + 56 \,\, s^3 - 4 \,\, \Lambda \,\, H}
{H \,\, s^2 + 16 \,\, s^3 - 4 \,\, \Lambda \,\, H} \right) \frac{H}{s} \,\,\, ,
\label{ORBITS}
\ee
obtained from Eqs.(\ref{phiprime}) and (\ref{hprime}). 
Figure 2 shows the solution
(\ref{sol1}) in the variables $s$ and $H$, which is represented by a solid line.

\vspace{0.5cm}
\begin{center}
\epsfig{file=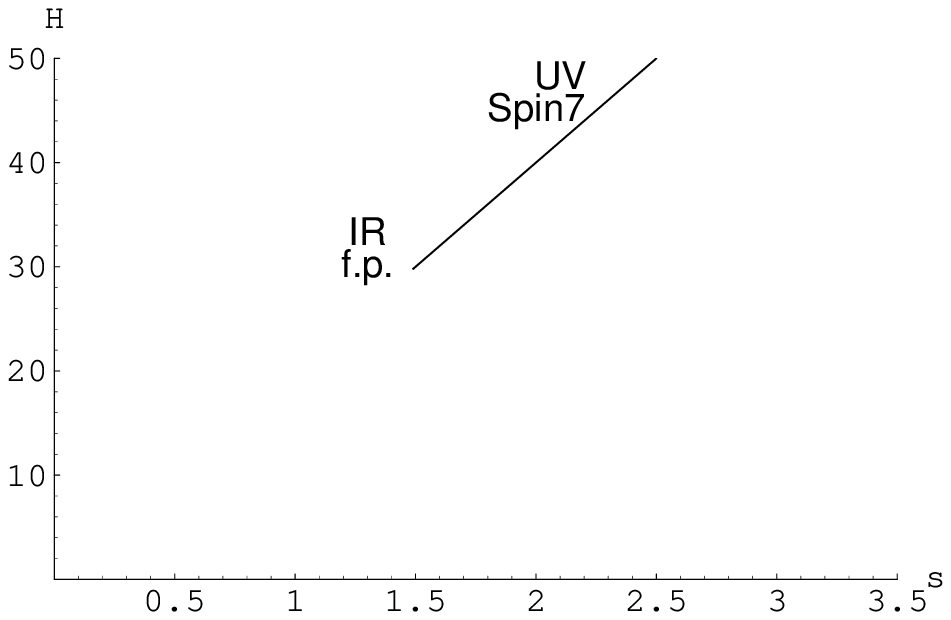, width=9cm}
\end{center}
\vspace{0.5cm}
\baselineskip=13pt
\centerline{\small{Figure 2: Solution as given by Eq.(17).
The vertical axis is $H(s)$ }}
\centerline{\small{and the horizontal one is $s$.}}

\vspace{0.5cm}

\baselineskip=15.5pt

If one approaches to the fixed point at the end of the semi--infinite
line $H=20 s$, one obtains that $f\rightarrow-\infty$ which confirms the IR character
of the fixed point. Note that in the figure we fixed $\Lambda=1$ and
one can easily see that as $\Lambda$ increases, the IR end-point moves
upwards along $H=20 s$. In Appendix C we show a more detailed plot of the
orbits of Eq.(\ref{ORBITS}).

The conclusion is that at UV the metric asymptotically corresponds to a three-dimensional
Minkowski spacetime times an Spin(7)
holonomy manifold. Then, it flows to the IR fixed point ($AdS_4 \times \tilde{S^7}$) 
along $H=20 s$ . 

One can consider more general solutions to the BPS equations
(\ref{fprime})--(\ref{hprime}) by turning on one more degree of
freedom, namely allowing also the difference $\phi(\tau)\,-\,h(\tau)$
in Eq.(\ref{sol1}) to be $\tau$-dependent. 
This solution and similar generalizations of
the solutions of next section are obtained in Appendix B. They
typically suffer from curvature singularities which render the field
theory interpretation difficult. However, we present some cases where
curvature singularities at IR are acceptable by Gubser's criterion
\cite{Gubser:2000nd}.

\section{From CY4 folds to AdS spaces and FT duals}
\subsection{D6-branes wrapping cycles inside CY3 folds}

Now, we will study a system in an analogous way as in the previous section,
but preserving twice the number of supersymmetries compared to
the above $E^{2,1} \times Spin(7)$ example.
Therefore, we will consider a D2-D6 system where the D6-branes
are wrapping a four-cycle of constant curvature which, in the present 
case, will be $S^2 \times S^2$. At the UV limit this 
four-cycle is inside a CY3 fold, and the
number of supercharges preserved by this configuration becomes 4, thus
leading to an ${\cal {N}}=2$ SYM theory in three dimensions. 
On the other hand, in the IR limit
we will obtain an eleven-dimensional metric
corresponding to M2-branes on the tip of cone based over $Q^{1,1,1}$,
an Einstein-Sasakian manifold. This represents the gravity dual of a 
three-dimensional ${\cal {N}}=2$ SCFT. 
In this section we want to argue that special holonomy
manifold is a cone over a CY3 fold. 
Furthermore, we will see that it
can be resolved in two possible ways, both of them
require to turn on an additional degree of freedom in the gauged
supergravity. First way of resolution preserves the zero curvature of
the configuration and leads to special holonomy manifolds
representing duals of three-dimensional ${\cal {N}}=2$ gauge 
theories with a mass gap, 
whereas the other way turns on a degree of freedom in the lower dimensional
supergravity leading to a 
configuration of the form $AdS_4 \times Y_7$ in M-theory. Here our
interest concentrates on the latter type of resolutions, 
that as in the previous section, leads to different physical effects.

Let us start by considering the theory with D2 and D6-branes. Since we
want to wrap the D6-branes in a four-cycle of the form $S^2 \times S^2$, we
have to choose an eight-dimensional metric of the form
\be
ds_{8}^2 = e^{2 f} \, dx_{1,2}^2 + dr^2 + 
e^{2 h} \, \left(d\theta_1^2 + \sin^2\theta_1 \, d\varphi_1^2 + 
d\theta_2^2 + \sin^2\theta_2 \, d\varphi_2^2\right) \,\,\, ,
\ee
together with an Abelian gauge field 
$A^{(3)}= \cos\theta_1 \, d\varphi_1 + \cos\theta_2 \, d\varphi_2 .$
It means that now the normal bundle of $S^2 \times S^2$ is $U(1)$. Therefore, in order 
to define Killing spinors
by means of twisting, we must break the $SU(2)$ group down to $U(1)$. This
is achieved by turning on the field $\lambda(r)$ in the eight-dimensional gauged supergravity.
Indeed, this field makes a distinction between the three directions of the
$S^3$ ``external'' to the branes system, thus leading to a symmetry breaking $SU(2)\to U(1)$. 
For this reason, in gauged supergravity we choose a ``vielbein'' of the form 
$L_\alpha^i= diag(e^{\lambda}, e^{\lambda}, e^{-2 \lambda}),$
that generates $T^{11} = T^{22}= e^{2\lambda}$, 
$T^{33} = e^{-4\lambda}$, 
as well as,
\be
P_\mu^{11}= P_\mu^{22}= \partial_\mu \lambda\,\,\, , \;\;\;\;\,\,\,\;\;\;
P_\mu^{33}=-2  \partial_\mu \lambda \,\,\, ,
\ee
\be
Q_\mu^{12}=- g \, A_\mu^{(3)} \, , \;\;\;\;\;\; Q_\mu^{13}=g \, \cosh(3\lambda) \, A_\mu^{(2)} 
\, , \;\;\;\;\;\; 
Q_\mu^{32}=g \, \sinh(3\lambda) \, A_\mu^{(1)} \,\,\, .
\ee
As before, we have a $G_4$ field  (in flat index notation)
\be
G_{x_1 x_2 x_3 r} =\Lm \, e^{-2\phi-4h}  \,\,\, .
\ee
After plugging it into the supersymmetric variations of fermions, it produces
the following BPS equations,
\be\label{fprime2}
f'= -\frac{1}{3} \, e^{\phi - 2 h - 2\lambda} +\frac{1}{24} \, 
e^{-\phi} \, (e^{-4 \lambda} + 2 e^{2\lambda})+\frac{\Lm}{2}e^{-4 h -\phi} \,\,\, ,
\ee
\be\label{phiprime2}
\phi'= -e^{\phi - 2 h - 2\lambda} +\frac{1}{8} \, e^{-\phi} \, (e^{-4 \lambda} + 2 \,  e^{2\lambda})-
\frac{\Lm}{2} \, e^{-4 h -\phi}           \,\,\, ,                                     
\ee                                        
\be\label{hprime2}
h'= \frac{2}{3 \, }e^{\phi - 2 h - 2\lambda} +
\frac{1}{24} \, e^{-\phi} \, (e^{-4 \lambda} + 2 e^{2\lambda})-
\frac{\Lm}{2} \, e^{-4 h -\phi}         \,\,\, ,                                       
\ee                                       
\be\label{lambdaprime}
\lambda'= \frac{2}{3} \, e^{\phi - 2 h - 2\lambda} -
\frac{1}{12}e^{-\phi} \, (-2 e^{-4 \lambda} + 2 e^{2\lambda})\,\,\, .
\ee
When $\Lambda=0$ this system coincides with the one in ref.\cite{Gomis:2001vg} 
\footnote{We thank the authors for clarifying a missprint in their original version.}.

Indeed, in this case ($\Lambda=0$), a  simple solution can
be easily found 
\be
\lambda=\frac{1}{6} \, \log(2),\;\;\; e^{h}=\frac{3r}{4\sqrt{2}}, \;\;\;\phi =
h -\frac{7}{6} \, \log(2), \,\,\,\,\,\, 3f=\phi \,\,\, ,
\label{singular}
\ee
that lifted to eleven dimensions, and after a suitable change of radial
variable, reads as 
\ba
ds_{11}^2 &=& dx_{1,2}^2 +  d\rho^2 + \nn \\
&& \frac{\rho^2}{8} \, \left(d\Omega_1^2 +
d\Omega_2^2 + d\Omega_3^2 +
\frac{1}{2} \, (d\psi + \cos\alpha \, d\beta - \cos\theta_1 \, d\phi_1 - \cos\theta_2
d\phi_2)^2 \right) \,\,\, .
\label{metricsing}
\ea
The metric has a singularity at $\rho=0$. We are mainly interested in
resolving the singularity by turning on M2-branes. However, as an
aside, let us consider other ways of resolution which do not render 
the field theory ending in a conformal point as follows. 
In gauged supergravity it can be done giving the field $\lambda$
a radial dependence. Indeed, 
by allowing $\lambda$ to be variable we can obtain a more general
solution that can be continued towards the IR. 
By defining a function $w$ and changing variables like
\beq
w(\rho)= \frac{3 \rho^4 + 8 \rho^2 +6 + C\rho^{-4}}{6 (\rho^2 +1)^2} \, , \;\;\;\;\;\;dr= d\rho
\left(\frac{\rho^2 }{16 w^{5/3}}\right)^{1/4} \,\,\, ,
\eeq
we have a solution that reads
\ba
e^{-6\lambda}&=&w(\rho), \;\;\;\;\;\;\;\; e^{2/3 \phi}= \frac{\rho \, w^{1/6}}{4} \, , \nn \\
e^{2h}&=&\frac{(\rho^2 + 1) \, \rho \, w^{1/6}}{16}, \;\;\;\;\; \;\;\; f=\frac{1}{3} \, \phi \,\,\,  .
\ea
This generates an eleven-dimensional metric as
\ba
ds_{11}^2&=& dx_{1,2}^2 + \frac{1}{w(\rho)} \, d\rho^2 +
\frac{\rho^2}{4} \, d\Omega_1^2 \nn \\
&&+\frac{(\rho^2 +1)}{4} \,
(d\Omega_2^2 + d\Omega_3^2) +
\frac{\rho^2 w(\rho)}{4}(d\psi + \cos\alpha \, d\beta - \cos\theta_1 \, d\phi_1 -
\cos\theta_2 \, d\phi_2)^2) \,\,\, , \nn \\
&&
\ea
corresponding to the metric for ${\bf C}^2$ 
bundle over $CP^1 \times CP^1$. In the case in which the integration
constant $C=0$ we recover the solution given in 
\cite{Gomis:2001vg,Cvetic:2000db}, 
in the case of non-zero $C$ this seems to be another possible resolution.

Now, let us turn back to our main interest, resolution by dynamically
turning on $F_4$. To this end, we will analyze a different kind of
solutions of the
above BPS equations, with a non-zero parameter $\Lambda$ and
study their holographic RG flows as in the previous section.

~

{\it A solution driving the flow from $E^{2,1} \times CY4$ to $AdS_4 \times Q^{1,1,1}$}

Since in this case there is an additional degree of freedom, a suitable change of
variable, as compared to Eq.(\ref{rt}), is 
\beq
dr = e^{\phi+4 \lambda} \, d\tau \,\,\, .
\label{change2}
\ee
A solution of the system (\ref{fprime2})-(\ref{lambdaprime})
is 
\ba
h(\tau)&=& \frac{1}{4} \, \log \left(\frac{2^{8/3} \, \Lambda}{3} + A \, e^{3 \tau/2}\right) 
\, , \;\;\;\;\;\;\;\;\;\; \phi(\tau)= -7/6 \, \log(2) + h(\tau) \, , \nn \\ 
f(\tau) &=& \frac{\tau}{2} - \frac{1}{4} \, \log \left(2^{8/3} \, \Lambda + 3 \, A \, 
e^{3\tau/2}\right)  \, , \;\;\;\;\;\;\;\;\;\;
\lambda = \frac{1}{6} \, \log(2) 
 \,\,\, ,
\label{solcy}
\ea
where $A$ is an integration constant. The corresponding eleven-dimensional metric is given by
\ba
ds^2_{11} &=& \frac{2^{7/9} \,\times\, 3^{1/6} \,\times\, e^{ \tau}}{(3 \, A \, e^{3 \tau/2}+ 
2^{8/3} \, \Lambda )^{2/3}} \, dx_{1,2}^2  \nn \\
&&+ \frac{1}{2^{2/9} \,\times\, 3^{1/3}} \, (3 \, A \, e^{3 \tau/2}+ 2^{8/3} \, \Lambda )^{1/3} \, 
d\tau^2 
+ \frac{2^{7/9}}{3^{1/3}} \,
 (3 \, A \, e^{3 \tau/2} + 2^{8/3} \, \Lambda )^{1/3} \, \times  \nn \\
&& \left( d\Omega^2_1 + d\Omega^2_2 + d\Omega^2_3 + \frac{1}{2} \, 
(d\psi + \cos\alpha \, d\beta - \cos\theta_1 d\phi_1 - \cos\theta_2 d\phi_2)^2 \right)
 \,\,\, .
\label{metricn2}
\ea
In flat indices the four-form field strength is
$F_{x_1 x_2 x_3 \tau}= \frac{2^{7/9} \,\times\, 3^{7/6}}{(3 \, A \, 
e^{3 \tau/2}+ 2^{8/3} \, \Lambda)^{7/6}},$
The solution in the $H$ and $s$ variables defined in the previous
section is shown in figure 3.

\vspace{0.5cm}
\begin{center}
\epsfig{file=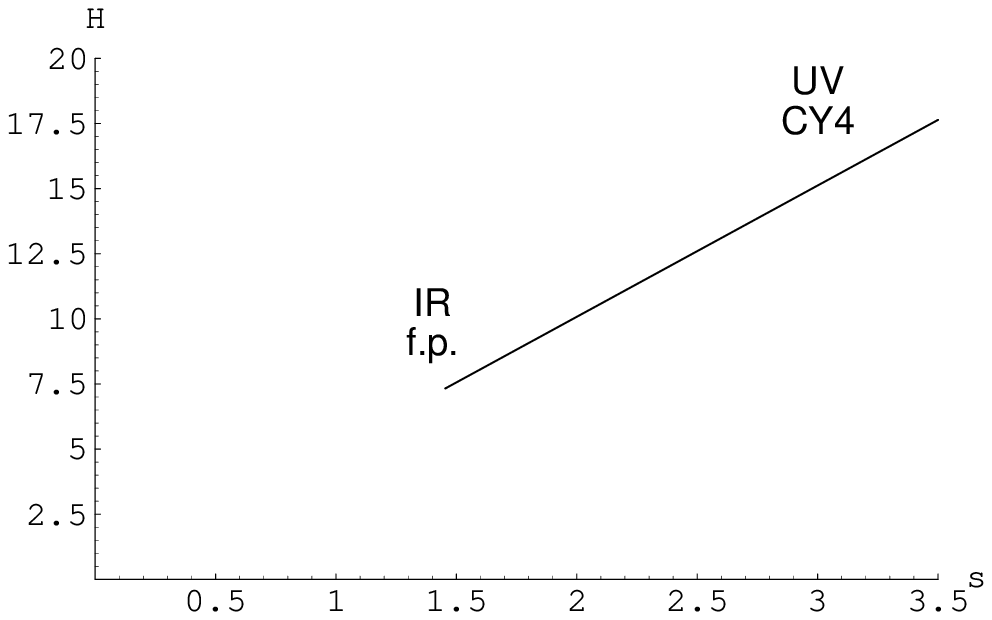, width=9cm}
\end{center}
\vspace{0.5cm}
\baselineskip=13pt
\centerline{\small{Figure 3: Solution as given by Eq.(39).
The vertical axis is $H(s)$ }}
\centerline{\small{and the horizontal one is $s$.}}

\vspace{0.5cm}

\baselineskip=15.5pt

Using the change of variables given in Eq.(\ref{change2}) it is easy to see
that the UV limit corresponds to $\tau \to +\infty$, while the IR one is for
$\tau \to -\infty$.
Therefore, let us first consider the limit $\tau \to \infty$. In M-theory, 
in the UV limit ($\tau\to +\infty$), the system
is turning off the $F_4$ field, such that the eleven-dimensional
configuration must be pure metric. 
The number of
preserved supercharges will be 4 and the metric will have an expression of
the form $E^{2,1} \times  CY4$, {\it i.e.} three flat dimensions plus
a non-compact CY4 fold. This metric reads \footnote {We rescaled the variable 
$\frac{e^{\tau/4}}{2^{1/9}} \, d\tau=d\rho$.}
\ba
ds_{11}^2 &=& dx_{1,2}^2 +  d\rho^2 + \nn \\
&& \frac{\rho^2}{8} \, \left(d\Omega_1^2 +
d\Omega_2^2 + d\Omega_3^2 +
\frac{1}{2} \, (d\psi + \cos\alpha \, d\beta - \cos\theta_1 \, d\phi_1 - \cos\theta_2
d\phi_2)^2 \right) \,\,\, .
\label{metricsing2}
\ea

It is singular at $\rho=0$. 
As discussed above resolution of this singularity is achieved by
flowing towards IR where $F_4$ is dynamically turned on. 
Hence, one should consider the complete metric
(\ref{metricn2}). This metric is the 
M-theory  dual of ${\cal {N}}=2$ SYM theory in three dimensions.
In three dimensions the ${\cal {N}}=2$ supersymmetric algebra is the reduction of the
${\cal {N}}=1$ in  four dimensions.
The role of the central charge is played by the four component of the
momentum. As in the higher dimensional case, the R-symmetry group is $U(1)_R$.

As usual,  scalars in the vector superfield parameterize the Coulomb
branch of the theory. In this case some fundamental hypermultiplets exist in the
Lagrangian which will describe the Higgs branch. 
In our case, we expect a dynamical scalar field coming from fluctuations of 
the metric. Another way of understanding the presence of this dynamical 
scalar is by noticing that D6-branes wrap an
$S^2 \times S^2$ cycle inside a  complex three-dimensional CY
space. Thus, there will be
one free direction (codimension one) in contrast to the $G_2$ case analyzed in the previous
section. This direction is interpreted as 
a scalar field describing the Coulomb branch of the theory.

In addition, due to the presence of D2-branes, or due to the $C_3$
field in M-theory, some other scalar modes will have
dynamics on the $2+1$-dimensional worldvolume.

In a  non-Abelian
theory, when we move into the Coulomb branch, we can dualize the vectors
to scalars in linear multiples $A_\mu \to \gamma$ and the Coulomb branch
is parametrized by the complex scalar $\Phi=\varphi+i\gamma$, like in
four dimensions, this is the factor that appears when the instanton
contributions are taken into account.
In the case in which we have fundamental hypermultiplets, the interaction term is
typically of the form
\beq
V\sim \int d^2\theta d^2\bar{\theta} \bar{Q} e^{V}Q \,\,\, .
\eeq
So, a bosonic term will be of the form $\bar{q}\varphi q$, this means
that in general, the Coulomb branch and the Higgs branch
will be disconnected. Nevertheless, there are situations in which we have
a mixing of Coulomb and Higgs branches. We will not discuss these cases
here, since our configurations will not have hypermultiplets. 
As it was studied by
Affleck, Harvey and Witten, the instantons that are associated with
$\Pi_2(G)$ are only present when we have a non-Abelian
gauge group and we consider the Coulomb branch, so $\Pi_2(G/U(1)^r)= Z^r$.
These instantons generate a superpotential 
that in the large $N$ limit goes to $W\sim e^{-N}$, so we cannot see it in a
supergravity approximation. This is the reason why a brane probe of our
configuration will lead to a 2-dimensional flat Coulomb branch.
It would be of much interest, 
to find dual gravity configurations to theories with fundamental hypermultiplets.

On the other limit of the flow, {\it i.e.} when $\tau\to-\infty$, one obtains the metric
\ba
ds_{11}^2 &=& e^{\xi \rho} \, dx_{1,2}^2 + d\rho^2 + 2 \, \left(\frac{4
|\Lm|}{3}\right)^{1/3} \times \nn \\
&& \left(d\Omega_1^2 + d\Omega_2^2 + d\Omega_3^2 + 
\frac{1}{2} \, (d\psi + \cos\alpha \, d\beta - \cos\theta_1 \, d\phi_1 - \cos\theta_2 
\, d\phi_2)^2 \right) \,\,\, ,
\label{q111}
\ea
where as before, $d\Omega_i^2$ denotes the line element over a
two-sphere\footnote{We rescaled the variable $\tau$ as
$\frac{2^{1/3} \, \Lambda^{1/6}}{3^{1/6}} \, d\tau=d\rho$.}.
In addition, one has a four-form field of the form
$F_{x_1 x_2 x_3 \tau}= \frac{2^{7/9} \, 3^{7/6}}{(2^{8/3} \, \Lambda)^{7/6}}$.
The manifold in Eq.(\ref{q111}) is $AdS_4 \times Q^{1,1,1}$ and the conformal
field theory to which this manifold is dual is well-known.
Indeed, the manifold $Q^{1,1,1}$ has been well studied in the past. The isometries of this 
space are $SU(2)^3 \times U(1)$ and are in correspondence with the global
symmetries of the CFT. The KK modes on $ Q^{1,1,1}$ were worked out in \cite{Merlatti:2000ed}.
It contains short and long multiplets of $Osp(2|4)$.

By linearizing the BPS equations
(\ref{phiprime2}\,-\,\ref{lambdaprime}) near the fixed point, one can work out the scale
dimensions of the operators dual to the eigenvectors of the mass matrix
which are combinations of $\phi$, $h$ and $\lambda$. Dimensions turn
out to be, 
\be
\Delta_{1}\,\,=\,\,6\, ,
\;\;\;\;\;\;\;\;\Delta_{2,\,3}\,\,=\,\,
\frac{3}{2}\,\pm\,\sqrt{\frac{31}{12}} \, .
\ee
Note, however, that in Eq.(\ref{solcy}) latter two of the eigenvectors are
turned off by requiring $\lambda\,=\,const.$ and
$h\,-\,\phi\,=\,const.$. Thus, as in the previous example, one 
is left with an irrelevant ``overall size'' deformation of $Q^{1,1,1}$ 
with dimension $\Delta\,=\,6$ and mass, $M^2\,=\,72$. Therefore the dual
CFT is at IR. Solutions obtained by turning on more eigenvectors are
presented in Appendix B, for which AdS geometry appears at UV, as one
expects. KK spectrum of $AdS_4\times Q^{1,1,1}$ compactification was
partially worked out in \cite{Fre:1999} and one indeed finds that the
``overall size'' deformation is a singlet of the bosonic isometry 
$SU(2)\times SU(2)\times SU(2)$ and $U(1)$ $R$-symmetry together
with (in our normalization conventions) $\Delta\,=\,6$ and mass, $M^2\,=\,72$.   

The CFT dual to the metric (\ref{q111}) corresponds to the one in an 
M2-brane on the tip of a cone on the seven-dimensional manifold \cite{
Fabbri:1999ag,Dall'Agata:1999hh,Ahn:1999ec,Oh:1998qi,Fabbri:1999ay,
Acharya:1998db,Morrison:1998cs,Herzog:2000rz}. This gauge
theory has a moduli space of vacua isomorphic to $Q^{1,1,1}$.

Like other CFT's the theory has a Coulomb branch described by fields in
the vector multiplet and a Higgs branch described by fields in chiral
multiplets. Working out the theory whose Higgs branch
is dual to the conifold above, one finds that fundamental
fields are doublets with respect to the flavour group $SU(2)^3$ : $A_i, B_i,
C_i$ with $i=1,2$, {\it i.e.}, the
fields transform as $A_i=(2, 1,1), \,\, B_i=(1, 2, 1), \,\, C_i=
(1, 1, 2)$ under the flavour group. The gauge theory has the color symmetry  
$SU(N) \times SU(N)\times SU(N)$, with elementary
degrees of freedom transforming
in the fundamental and anti-fundamental representation of the $SU(N)'s$, namely 
$A_i=(N, \bar{N},1), B_i=(1, N, \bar{N}), C_i= (\bar{N},
1, N)$. 
These fields have conformal weight $c=1/3$, hence one can construct gauge
invariant operators of the form
\beq
X^{ijk}=A^i B^j C^k \,\,\, ,
\eeq
out of them. 
These eight operators are singlets under the global symmetries and have
conformal weight equal to one.

One important point is the comparison of the KK modes on the
$Q^{1,1,1}$ manifold and the spectrum of hypermultiplets in the CFT. The
spectrum of the Laplacian in $Q^{1,1,1}$ is computed and one can
associate it with 
a chiral multiplet in the $(k/2,k/2,k/2)$ representation of $SU(2)^3$
with dimension $E=k$. Therefore, it is natural to make a correspondence with
composite operators of the form $Tr(ABC)^k$ with symmetrized $SU(2)$ indices.
This agrees with the gauge theory.

Nevertheless, there are some operators in gauge theory--like those where
the $SU(2)$ indices are not symmetrized--that do
not have a KK analog. One would think that a superpotential can be
generated in such a way to get rid of those states as in the case of
$T^{1,1}$, but this is not the case. Indeed, we can see from a gauge theory
perspective that the potential that should do the job
\ba
V &\sim& [(|A_1|^2 +|A_2|^2 -|C_1|^2 -|C_2|^2)^2 +( |B_1|^2 +|B_2|^2
-|A_1|^2 -|A_2|^2 +)^2 \nn \\
&&+ (|C_1|^2 +|C_2|^2 -|B_1|^2 -|B_2|^2 )^2] \,\,\, ,
\ea
vanishes due to the fact that it is exactly the (Higgs branch)
description of the manifold $Q^{1,1,1}$. So, the potential does not solve
the problem and one needs to assume that these unwanted colored degrees of
freedom are not chiral primaries.

One can also find the presence of a baryonic operator essentially corresponding to
wrapping an M5-brane on a five-cycle inside the eight-cone.
The operators corresponding to baryons are of the form $det[A], \,\, det[B], \,\,
det[C]$. Since our manifold $Q^{1,1,1}$ has Betti numbers $b_2, b_5$
different from zero, there is another $U(1)$ under which only
non-perturbative states will be charged. In our case, the baryonic symmetry
acts on the fundamental fields as $A_i=(1, -1,0), \,\, B_i=(0, 1, -1), \,\, C_i=
(-1, 0, 1)$, therefore we can see that gauge invariant operators $X$
are not charged under baryon number. One can compute the dimension of the
baryonic operator by computing the mass of an M5-brane wrapping a
five-cycle inside the cone. This mass in the case of a supersymmetric cycle,
coincides with the volume of the cycle. In our case the five-cycle is a $U(1)$
fibre over $S^2 \times S^2$ and since our manifold is a $U(1)\to S^2
\times S^2 \times S^2$ we have three different supersymmetric cycles that
are associated with the three operators defined above. Each cycle is
supersymmetric as we can see from the twisting condition described
above. The  volume of the cycle, can be computed to be proportional to $N/3$ thus
confirming the fact that each operator $A,B, C$ have dimension $1/3$. 
If the M5-brane wraps a three-cycle, the object is
interpreted as a domain wall of the CFT.

\subsection{The case of ${\cal {N}}=3$ supersymmetry: from HK to $N^{0,1,0}$}

Here, we will briefly comment on the case where the D6-branes are wrapping
a four-cycle that preserves ${\cal {N}}=3$ supersymmetry. The set up is very similar to
the previous examples, except we take the four-cycle to be
a $CP^2$ manifold. We choose a metric (using as 
coordinates $\xi$ and the three angles in the left-invariant forms 
$\sigma^i$)
\beq
ds_{CP2}^2 = d\xi^2 + \frac{1}{4} \, \sin^2\xi \, (\sigma_1^2 + \sigma_2^2 +
cos^2\xi \, \sigma_3^2) \,\,\, .
\eeq
The gauge field which provides the twisting preserving six
supercharges is given by
\beq
A^{(i)}= \cos\xi \, \sigma^{(i)}, \,\,\, \;\; A^{(3)}= \frac{1}{2} \, (1 + \cos^2\xi)
\sigma^{(3)} \,\,\,.
\eeq
A solution of the BPS equations lifted to eleven dimensions reads
\beq
ds_{11}^2 = \frac{ dx_{1,2}^2}{\left(1 + \frac{B}{r^6}\right)^{2/3}} +
\left(r^6 + B\right)^{1/3} \, ds_{CP2}^2 +
2 \left(\frac{B}{r^6} +1\right)^{1/3} dr^2 + \frac{r^2}{2} \left(\frac{B}{r^6} +1\right)^{1/3} 
(\omega^i - A^i)^2  
\label{n010}
\eeq
together with the four-form field strength
\beq
F_{xyzr}= \frac{3 B}{(B + r^6)^{7/6}} \,\,\, ,
\eeq
written in flat indices, where $B$ is a constant.
Therefore, the metric (\ref{n010}) represents a holographic RG flow
from $E^{2,1} \times HK$ (hyperKahler) at UV to $AdS_4 \times N^{0,1,0}$ at IR.
The isometry group of $N^{0,1,0}$ is $SU(3) \times SU(2)$ 
while its holonomy is $SU(2)$.

As it is known, an ${\cal {N}}=3$ supersymmetric gauge theory has the field content of an ${\cal {N}}=4$
supersymmetric gauge theory, plus an interaction that respects three out of the four spinors.

It was shown by Kapustin and Strassler \cite{Kapustin:1999ha} that
for the Abelian case, the ways of breaking ${\cal {N}}=4$ down to  ${\cal {N}}=3$ supersymmetry 
are either by adding a Chern Simons term or with a mass term for a chiral superfield $Y^I$
in the adjoint representation of the gauge group.

In our case, we have a supergravity solution of the form $AdS_4 \times N^{0,1,0}$. 
The dual gauge theory in the IR limit will have a gauge group $SU(N) \times
SU(N)$ and a flavor group $SU(3)$. There will be two hypermultiplets, $u_1$, $u_2$ and
$v_1$, $v_2$ transforming in
the $(3, N, \bar{N})$ and $(\bar{3}, \bar{N}, N)$ representations and two chiral multiplets,
$Y_{(1)}, Y_{(2)}$ in the adjoint representation of $SU(N)$. There is a superpotential of
the form
\beq
V\sim \, g_i \, Tr( Y_{(i)} \, \vec{u} \, \cdot \, \vec{v} ) + 
\alpha_i \, Tr( Y_{(i)} \, Y_{(i)} ) \,\,\, ,
\eeq
where $g_i$ are the gauge couplings of each $SU(N)$ group and $\alpha_i$
are the Chern Simons coefficients. Interesting aspects of this theory,
like the KK spectrum of the compactifications and different checks of the
duality have been studied in \cite{Termonia:1999cs,Billo:2000zr}.

\section{Penrose limits and pp-waves}

In this section we will show how to obtain the pp-waves in the
Penrose limit for the IR region of the supergravity solutions
described in sections 4 and 5. 

We will focus on the solutions with ${\cal {N}}=2$ and ${\cal {N}}=3$ supersymmetry 
and we will add a brief description of the ${\cal {N}}=1$
case near the end of the section.

The interest of taking the Penrose limit is based on the fact that it could be possible to
define following \cite{Berenstein:2002jq} a Matrix model to check the correlation 
between the gravity and the gauge theory side.
We postpone the checks of this correlations 
for a future publication, here we will only 
concentrate on the gravity aspects. 

~

{\it Penrose limit of the AdS $\times$ Einstein-Sasakian manifold}

The Einstein metric of $AdS_4 \times Q^{1,1,1}$ can be written as
\be
ds^2_{11} = ds^2_{AdS_4} + ds^2_{Q^{1,1,1}} \,\,\, ,
\ee
where
\be
ds^2_{AdS_4} = R^2 \, (-dt^2 \, \cosh^2\rho + d\rho^2 + \sinh^2\rho \, d\Omega^2_2) \,\,\, ,
\ee
and
\bea
ds^2_{Q^{1,1,1}} &=& \mu^2 \, R^2 \, (d\theta_1^2 + \sin^2\theta_1 \, d\phi^2_1 + 
d\theta_2^2 + \sin^2\theta_2 \, d\phi^2_2 + d\theta_3^2 + \sin^2\theta_3 \, d\phi^2_3 + \nn \\
&& \frac{1}{2} \, (d\psi + \cos\theta_1 \, d\phi_1 + \cos\theta_2 \, d\phi_2 +
\cos\theta_3 \, d\phi_3)^2) \,\,\, .
\eea
Where $\mu$ is the relation between the radii of $AdS_4$ and $Q^{1,1,1}$. Topologically
$Q^{1,1,1}$ is a $U(1)$ bundle over $S^2 \times S^2 \times S^2$, so that 
it can be parametrized by $(\theta_1, \phi_1)$, $(\theta_2, \phi_2)$ and 
$(\theta_3, \phi_3)$ coordinates over each $S^2$, respectively, while the period of the 
Hopf fiber coordinate $\psi$ is $4 \pi$. The 
$SU(2)_1 \times SU(2)_2 \times SU(2)_3 \times U(1)$ isometry of $Q^{1,1,1}$ is identified with
the $SU(2)_1 \times SU(2)_2 \times SU(2)_3$ global symmetry and $U(1)_R$ symmetry of the
dual $SU(N)$ ${\cal {N}}=2$ SCFT in three dimensions.

Now, the idea is to obtain a certain scaling limit around a null geodesic in
$AdS_4 \times Q^{1,1,1}$. This rotates the $\psi$ coordinate of $Q^{1,1,1}$ in
correspondence with the $U(1)_R$ symmetry of the dual SCFT. Moreover, the changes in the
angles $\phi_1$, $\phi_2$ and $\phi_3$ generate an $U(1)_1 \times U(1)_2 \times U(1)_3 
\, \subset \, SU(2)_1 \times SU(2)_2 \times SU(2)_3$ isometry. 
In the SCFT side this is generated by the dual Abelian charges $Q_1$, $Q_2$ and $Q_3$,
which are the Cartan generators of the global  $SU(2)_1 \times SU(2)_2 \times SU(2)_3$ 
symmetry group of the field theory.

Thus we define new coordinates 
\bea
x^+ &=& \frac{1}{2} \, \left(t + \frac{\mu}{\sqrt{2}} (\psi+\phi_1+\phi_2+\phi_3)\right) \,\,\, , \\
x^- &=& \frac{R^2}{2} \, \left(t - \frac{\mu}{\sqrt{2}} (\psi+\phi_1+\phi_2+\phi_3)\right) \,\,\, . 
\eea
Note the scaling in the latter equation by $R^2$. We will consider a scaling limit around
$\rho=\theta_1=\theta_2=\theta_3=0$ in the metric above, such that when we take the
limit $R \to \infty$ we also scale the coordinates as follows
\be
\rho=\frac{r}{R} , \,\,\,\,\,\, \theta_1=\frac{\zeta_1}{R} , \,\,\,\,\,\, 
\theta_2=\frac{\zeta_2}{R} , \,\,\,\,\,\, \theta_3=\frac{\zeta_2}{R} .
\ee
Therefore, the Penrose limit of the $AdS_4 \times Q^{1,1,1}$ metric is given by
\be
ds^2_{11} = -4 \, dx^+ \, dx^- + \sum^3_{i=1} \, (dr^i \, dr^i - 
r^i \, r^i \, dx^+ \, dx^+) +
  \sum^3_{i=1} \, (\mu^2 d\zeta^2_i+\mu^2 \zeta_i^2 \, d\phi^2_i-
\frac{\mu}{\sqrt{2}} \, \zeta_i^2 \, d\phi_i \, dx^+) \,\,\, .
\label{metricapenrose}
\ee
Changing to the complex coordinates $z_j=\zeta_j \, e^{i \phi_j}$ one obtains
\be
ds^2_{11} = -4 \, dx^+ \, dx^- + \sum^3_{i=1} \, (dr^i \, dr^i - 
r^i \, r^i \, dx^+ \, dx^+) + 
 \sum^3_{j=1} \, 
(\mu^2 dz_j \, d{\bar{z}}_j + i \frac{\mu}{2 \sqrt{2}}\, ({\bar{z}_j} \, dz_j - z_j \, d{\bar{z}_j} ) \, 
dx^+ ) \,\,\, .
\ee
This metric has a covariantly constant null Killing vector $\partial / \partial x^{-}$, and
therefore is a pp-wave metric having a decomposition of  
${\bf R}^9$ as ${\bf R}^3 \times {\bf R}^2 \times {\bf R}^2 \times {\bf R}^2$.
Three-dimensional Euclidean space is parametrized by $r_i$, while 
${\bf R}^2 \times {\bf R}^2 \times {\bf R}^2$ is parametrized by $z_j$
above. In addition, the background has a constant $F_{+ x_1 x_2 r}$. The symmetries of this 
configuration are the $SO(3)$ rotations in ${\bf R}^3$ and 
$U(1) \times U(1) \times U(1)$ symmetry related to the 
${\bf R}^2 \times {\bf R}^2 \times {\bf R}^2$ rotations. 
We choose this particular Penrose limit due to the fact that these
$U(1)$'s are representing the symmetries of the gauge theory
dual to this background.
From  the dual field theory viewpoint, the $SO(3)$ isometry  is a subgroup
of the $SO(2,3)$ conformal group. $U(1) \times U(1) \times U(1)$
rotational charges $J_1$, $J_2$ and $J_3$ correspond to
differences between $U(1) \times U(1) \times U(1)$ charges $Q_1$, $Q_2$
and $Q_3$ and the $U(1)_R$ charge. Indeed from the field theory side,
it is expected to deal with operators with large $U(1)_R$
symmetry charge $J$, being the charges  $Q_1$, $Q_2$ and $Q_3$
also scaled as $\sqrt{N}$ like $J$,
while the corresponding rotational charges $J_i$'s would remain finite.

After an  $U(1) \times U(1) \times U(1)$ rotation  in the 
${\bf R}^2 \times {\bf R}^2 \times {\bf R}^2$ plane as
\be
z_j = e^{i \, \sqrt{2} \, x^+/(4 \, \mu)} \, w_j , \,\,\,\,\,\,\,\,\, 
{\bar {z}}_j = e^{-i \, \sqrt{2} \, x^+/(4 \, \mu)} \, {\bar{w}}_j , 
\label{resc}\ee
one obtains a metric that after suitable rescalings of its variables turns out to be
\be
ds^2_{11}=-4 \, dx^+ \, dx^- - \left(\left(\frac{\mu}{6}\right)^2 \,
{\vec{r}}_3^{\, 2} + \left(\frac{\mu}{3}\right)^2 \, {\vec{y}}_6^{\, 2} \right) \, dx^+ \, dx^+ + 
d{\vec{r}}_3^{\, 2} + d{\vec{y}}_6^{\, 2} \,\,\, ,
\ee
where $y_j$ are the 6 coordinates on ${\bf R}^2 \times {\bf R}^2 \times {\bf R}^2$.
The above metric corresponds to the maximally supersymmetric pp-wave solution of $AdS_4 \times S^7$. 
It means that the dual SCFT is ${\cal {N}}=8$, $SU(N)$ super
Yang Mills theory in three dimensions. This shows the enhancement of
supersymmetry analogous to the ones obtained in
\cite{Itzhaki:2002kh,Gomis:2002km,Zayas:2002rx}.
This fact might be interpreted as a hidden ${\cal {N}}=8$ supersymmetry
which was already present in the corresponding subsector of the dual ${\cal {N}}=2$ SCFT.

~

{\it Penrose limit of the AdS $\times N^{0,1,0}$ manifold}

The Einstein metric of 
$AdS_4 \times N^{0,1,0}$ can be written as
\be
ds^2_{11} = ds^2_{AdS_4} + ds^2_{N^{0,1,0}} \,\,\, ,
\ee
where we again use
\be
ds^2_{AdS_4} = R^2 \, (-dt^2 \, \cosh^2\rho + d\rho^2 + \sinh^2\rho \, d\Omega^2_2) \,\,\, ,
\ee
while
\bea
ds^2_{N^{0,1,0}} &=& \mu^2 \, R^2 \, (d\zeta^2 + \frac{\sin^2\zeta}{4} \, (d\theta^2 + 
\sin^2\theta \, d\phi^2 + \cos^2\zeta \, (d\psi + \cos\theta \, d\phi)^2) + \nn \\
&& \frac{1}{2} \, (\cos\gamma \, d\alpha + \sin\gamma \, \sin\alpha \, d\beta -
\cos\zeta \, (\cos\psi \, d\theta + \sin\psi \, \sin\theta \, d\phi))^2 + \nn \\
&& \frac{1}{2} \, (-\sin\gamma \, d\alpha + \cos\gamma \, \sin\alpha \, d\beta -
\cos\zeta \, (-\sin\psi \, d\theta + \cos\psi \, \sin\theta \, d\phi))^2 + \nn \\
&& \frac{1}{2} \, (d\gamma + \cos\alpha \, d\beta - \frac{1}{2} \,
(1+\cos^2\zeta) \, (d\psi + \cos\theta \, d\phi))^2 ) \,\,\, .
\eea
Where $\mu$ is the relation between the $AdS_4$ and 
$N^{0,1,0}$ radii.
Again, the idea is to obtain a certain scaling limit around a null geodesic in
$AdS_4 \times N^{0,1,0}$. In this case we can define the coordinates 
\bea
x^+ &=& \frac{1}{2} \, \left(t + \frac{\mu}{\sqrt{2}} (\gamma+\beta-\psi/2-\phi/2)\right) \,\,\, , \\
x^- &=& \frac{R^2}{2} \, \left(t - \frac{\mu}{\sqrt{2}} (\gamma+\beta-\psi/2-\phi/2)\right) \,\,\, .
\eea
We will consider a scaling limit around
$\rho=\theta=\alpha=0$ and $\zeta=\pi/2$ in the metric above, so that when we take the
limit $R \to \infty$ we also scale the coordinates as
\be
\rho=\frac{r}{R} , \,\,\,\,\,\, \zeta=\frac{\pi}{2} + \frac{x}{R} , \,\,\,\,\,\, 
\alpha=\frac{z}{R} , \,\,\,\,\,\, \theta=\frac{y}{R} .
\ee
Therefore, after some appropriate redefinitions of coordinates,
the Penrose limit of the $AdS_4 \times N^{0,1,0}$ metric is given by
\bea
ds^2_{11} &=& -4 \, dx^+ \, dx^- + \sum^3_{i=1} \, (dr^i \, dr^i - 
r^i \, r^i \, dx^+ \, dx^+) + \mu^2 \, (\, z^2 \, d\beta^2 + \, y^2 \, d\phi^2 
+ x^2 \, d{\hat{\psi}}^2 ) +\nn \\
&& \mu^2 \, (dx^2 + dy^2 + dz^2)
- \sqrt{2} \, \mu \, dx^+ \, (\, z^2 \, d\beta + \, y^2 \, d\phi + x^2 \, d{\hat{\psi}})  \, ,
\eea
where we have changed $\psi+\phi \to {\hat{\psi}}$. 

Using the redefinitions 
\be
x=\zeta_1 , \,\,\,\,\,\,\, y=\zeta_2 , \,\,\,\,\,\,\, z=\zeta_3 , 
\ee
and 
\be
{\hat{\psi}}=\phi_1 , \,\,\,\,\,\,\, \phi=\phi_2 , \,\,\,\,\,\,\, \beta=\phi_3 , 
\ee
together with a further rescaling, this metric becomes exactly 
Eq.(\ref{metricapenrose}). Therefore, we can follow a similar path
defining complex coordinates, etc., obtaining that the Penrose limit of
$AdS_4 \times N^{0,1,0}$ reduces to the same pp-wave as the corresponding one
of $AdS_4 \times S^7$. Hence, we might conclude that likely there is 
a hidden ${\cal {N}}=8$ supersymmetry
which was already present in the corresponding 
subsector of the dual ${\cal {N}}=3$ SCFT.

~

{\it Penrose limit of the AdS $\times$ squashed seven-sphere}

Finally, we would like to add some comments on the case of the Spin(7) holonomy
manifold, that is our solution preserving ${\cal {N}}=1$ supersymmetry.
This case is analogous to the Penrose limit for the gravity solution
of D5-branes on the resolved conifold, that has been worked out in 
\cite{Gomis:2002km} \footnote{We thank Jaume Gomis for explanations on
this respect.}.

The Einstein metric of 
$AdS_4 \times {\tilde{S}}^7$ can be written as
\be
ds^2_{11} = ds^2_{AdS_4} + ds^2_{{\tilde{S}}^7} \,\,\, ,
\ee
where as before we have
\be
ds^2_{AdS_4} = R^2 \, (-dt^2 \, \cosh^2\rho + d\rho^2 + \sinh^2\rho \, d\Omega^2_2) \,\,\, ,
\ee
while 
\be
ds^2_{{\tilde{S}}^7} = \mu^2 \, \left( d\Omega^2_4 + \frac{1}{5} \, 
(\omega^i-A^i) \right) \,\,\, ,
\ee
where $\mu$ stands for the relation between AdS and squashed seven-sphere radii, and
the factor $1/5$ comes from the relation between $S^4$ and $SU(2)$-group manifold radii.
We remind that the isometries of ${\tilde{S}}^7$ are $SO(5) \times SU(2)$.

Indeed, we can proceed in the following way, we rescale the coordinates such that 
the part coming from the four-sphere reads 
\beq
d\Omega_4^2 \approx  \frac{d\tau^2}{R^2 } + \frac{\tau^2}{4 R^2} \, d\Omega_3^2 \,\,\, ,
\label{mb}
\eeq
that is an ${\bf R}^4$ space. This rescaling includes $\rho \to r/R$,
$\theta \to y/R$ and $\alpha \to z/R$, where the angles are used to define
the left-invariant one forms and the precise definition is given in Appendix D.

In this coordinates the gauge field will be approximated by
$A^i \approx (1 - \frac{\tau^2}{R^2 }) \, \sigma^i$. 
In the limit of large $R$, the term
in the metric describing the fibration between the coordinates of the three-sphere 
and the four-sphere will basically consist of two parts. 
After a suitable change of variables,
the term coming from $(\omega^{i} -\sigma^{i})^2$ will contribute to   
$(dx^+ - dx^-/R^2)^2$ and a flat two-dimensional space.  
The second term in the metric (\ref{mb}) proportional to $\tau$ will contribute with a term of the form
$\tau^2/R^2 dx^+ d\phi$. After a similar 
rescaling as the one for the ${\cal {N}}=2$ case is done, it will add a mass
term for two of the flat directions, and we obtain a metric that looks very similar to
Eq.(\ref{metricapenrose}).

\section{RG flows from gauged supergravity in $6$ dimensions}

In this section we will present other Holographic RG flow examples 
obtained from $F(4)$ gauged supergravity \cite{romans6} in 6 dimensions. These
will be of interest since uplifting of these solutions to massive IIA supergravity
is known \cite{cveticmassive}. The bosonic Lagrangian is
\be
e^{-1} \, {\cal{L}}^{(6)}_B = - \frac{1}{4} \, R + (\partial^\mu \varphi)
(\partial_\mu \varphi) - V(\varphi) \,\,\, ,
\label{lagrangian}
\ee
where we set the Abelian, non-Abelian and the two-index tensor gauge
fields to zero. The dilaton potential is given by
\be
V(\varphi) = - \frac{1}{8} \, (g^2 \, e^{2 \varphi} + 4 \, m \, g \, e^{-2 \varphi}
             - m^2 \, e^{-6 \varphi}) \,\,\, ,
\label{dilaton-potential}
\ee
where $g$ is the non-Abelian coupling constant and $m$ becomes the mass
of $B_{\mu \nu}$ field via Higgs mechanism \cite{romans6}. Figure 4 shows
the dilaton potential.
Without loss of generality one can set $g=3m$ which will yield a
supersymmetric background at the maximum of $V$ at $\varphi=0$. 
At the two extrema of the potential,
\be
\varphi_{min} = - \frac{1}{4} \, \log(3), \,\,\,\, V(\varphi_{min}) 
                  = - \frac{3 \, \sqrt{3}}{2} \, m^2 \,\,\, ,
\label{bcmin}
\ee
and,
\be
\varphi_{Max} = 0 , \,\,\,\, V(\varphi_{Max}) 
                  = - \frac{5}{2} \, m^2 \,\,\, .
\label{bcmax}
\ee
which correspond to $AdS_6$ solutions with curvatures  
$R_{Min}= 9 \, \sqrt{3} \, m^2$ and  $R_{Max} = 15\, m^2$, 
respectively.   
The Euler-Lagrange equations are
\ba
0 &=& R_{\mu \nu} - 4 \partial_\mu\varphi \, \partial_\nu\varphi + g_{\mu \nu} 
      \, V(\varphi) \,\,\, , 
\label{Einstein-eq} \\ 
0 &=& - 2 \Box \varphi  - \frac{\partial V}{\partial \varphi} \,\,\, .
\label{dilaton-eq}
\ea

These fixed-point solutions can be lifted to massive type IIA
string theory using the uplifting procedure given by \cite{cveticmassive} 
where the Romans' F(4) gauged supergravity in 6 dimensions is obtained
from a consistent warped $S^4$ reduction of massive type IIA string theory.

\vspace{0.5cm}
\begin{center}
\epsfig{file=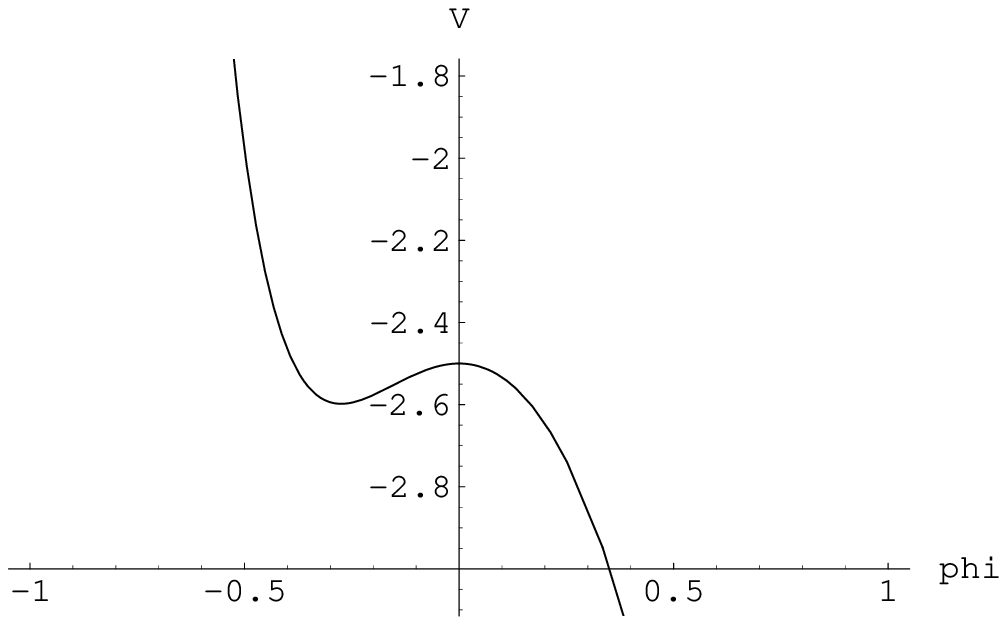, width=9cm}
\end{center}
\vspace{0.5cm}
\baselineskip=13pt
\centerline{\small{Figure 4: Scalar potential for 6-dimensional}}
\centerline{\small{gauged supergravity in units $m^2=1$.}}

\vspace{0.5cm}

\baselineskip=15.5pt


\subsection{Interpolating Solutions}

{\it Non-supersymmetric flow}

One can obtain a kink solution to EOM which interpolates between local maximum and minimum of 
$V(\phi)$. We make the usual domain-wall ansatz, 
\be
ds^2=e^{2 f(r)} \, \eta_{\mu \nu} dx^\mu x^\nu - dr^2 \,\,\, ,
\ee
with mostly minus convention. For $f(r)=r/l$ the metric becomes $AdS_6$ with $l$ constant. The
kink is the general solution $f(r)$ interpolating between two $AdS_6$ spacetimes 
with different radius $l$.
In these coordinates the UV limit is given when $f \rightarrow +\infty$, while the
IR limit corresponds to $f \rightarrow -\infty$. 
One obtains the following EOM from 
Eq.(\ref{Einstein-eq}) and Eq.(\ref{dilaton-eq}).   
\ba 
A'' & = & - (\varphi')^2 \,\,\, ,\\
5 \, A' \, \varphi' + \varphi'' &=& +\frac{1}{16} \, \frac{\partial V(\varphi)}{\partial \varphi} \, ,
\ea
with the following boundary conditions
\begin{equation}
\varphi'\big|_{\phi=0}=\varphi'\big|_{\phi=-\frac{1}{4}}=0 \,\,\, .
\end{equation}
It is easy to see that a superpotential $W(\phi)$ defined 
by,
\be\label{spot}
-V(\varphi) = 5 W(\varphi)^2 -\left(\frac{\pt W}{\pt \varphi}\right)^2,
\ee
exists but does not have two extrema. Hence the kink interpolating
between two $AdS_6$ solutions will necessarily be non-supersymmetric. In
this case, one has to solve the second order EOM given above. The
solution corresponds to flowing towards left from the origin in
figure 4.  
We could not find an analytic solution mostly due to the lack of supersymmetry along the flow,
however a numerical solution can be obtained. A similar study of an RG
flow from seven-supergravity was done in \cite{Campos:2000}. 
Figure 5 shows the interpolating solution. 

In order to identify the dual field theories at both UV and IR ends of the
flows, one solves the scalar equation of motion linearized near the
extrema. The solution near the UV fixed point reads, 
\be\label{phiUVexp}
\varphi = A_1 e^{-2r} +  A_2 e^{-3r} \,\,\, .
\ee
Noting that $\varphi<0$ along the flow, we see that any VEV type
deformation is excluded, otherwise VEV's of the operators would be 
negative which is not physical. Therefore, the deformation is  
a source term with scale dimension $\De_1=3$ or $\De_2=2$. The second 
case is also excluded since there does not exist any bosonic
operator in five-dimensional CFT of dimension 2, hence we conclude that the RG flow
is initiated by deforming the CFT with a mass term to the chiral 
superfield, 
$$\int d^5 x \,\, Tr[\Phi\Phi] \,\,\, ,$$
with the scale dimension $\De_1=3$. In Eq.(\ref{phiUVexp}) this 
corresponds to setting $A_2=0$. 

\vspace{0.5cm}
\begin{center}
\epsfig{file=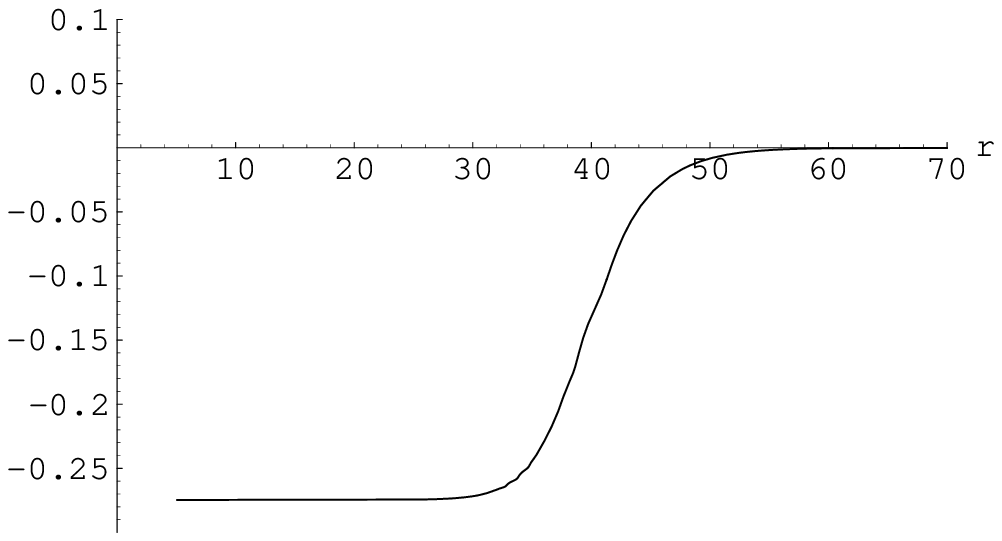, width=9cm}
\end{center}
\vspace{0.5cm}
\baselineskip=13pt
\centerline{\small{Figure 5: $\varphi$ as a function of $r$.}}

\vspace{0.5cm}

\baselineskip=15.5pt

Linearization near IR fixed point yields scale dimensions,
$$\De_{1,2} = \frac{1}{2}(5\pm\sqrt{65})$$
which shows that the source term acquires an anomalous dimension. 
This is of course expected in the absence of symmetries which would protect the scale
dimension if the operator.    
 
~

{\it Supersymmetric flow}

Another possible kink solution is interpolating between 
the maximum at $\varphi=0$ to $\varphi = +\infty$. This corresponds to 
flowing from the origin towards right in  figure 4. Since there is a 
curvature singularity at $\phi=+\infty$, one has to decide whether the flow can be physical 
by means of  the Gubser's criterion \cite{Gubser:2000nd}. Since the scalar potential is 
bounded from above in that limit we find that the curvature
singularity is good type.

A superpotential is obtained from, Eq.(\ref{spot}), (up to a sign), 
\be\lb{spot2}
W= \frac{1}{4}\left(3e^{\varphi}+e^{-3\varphi}\right).
\ee
Note that we are using limits in which $l$ is set to 1. 
Accordingly, second order EOM are reduced to the first order Killing spinor 
equation,
\be\lb{KSE}
\frac{\pt\varphi}{\pt r}=\frac{3}{4}\left(e^{-3\varphi}-e^{\varphi}\right) \,\,\, ,
\ee
with the solution,
\be\lb{sol6D}
r = const + \frac{1}{3}\left(2\arctan(e^{-\varphi})
-\log(1-e^{-\varphi})+\log(e^{-\varphi}+1)\right) \,\,\, .
\ee
Although one can not invert this equation into the form $\varphi(r)$, it
contains the same information. Note that in the UV limit $\varphi\to 0$,
$r\to +\infty$ and in the IR limit $\varphi\to +\infty$, $r\to const$ as
expected from an RG flow between conformal and non--conformal
theories \cite{zaffaroni}.   

Expansion of $W(\varphi)$ near the UV fixed point leads to the linearized
solution around the maximum, 
\be
\varphi= A\,\,e^{-3r} \,\,\, .
\ee 

Since there can not be a bosonic source term of dimension 1, we
conclude that this is a pure Higgs type deformation \cite{freedman2} 
where the $\De=3$  operator,  $m^2 Tr[\Phi\Phi]$ acquires a VEV. As a
result, conformal supercharges are broken whereas Poincare
supercharges are preserved, {\it i.e.} there are 8 supercharges along the
flow. In the dual field theory, $R$--symmetry 
group $SO(4)$ is broken down to
$SO(2)\times SO(2)$. This is the analog of Coulomb branch flow 
of ${\cal {N}}=4$ SYM \cite{freedman3}.

\section{Final comments}

We would like to summarize the different points studied in this paper.
We constructed a set of solutions describing D2-D6 brane system where
the D6-branes wrap different supersymmetry four-cycles in
several manifolds with special holonomy. 
In order to find these solution we have used the Salam-Sezgin eight-dimensional
gauged supergravity. These solutions represent holographic RG flows
for three-dimensional supersymmetric gauge field theories.
We have analyzed some aspects of the dual gauge theories
that turn out to be SCFT's preserving  ${\cal {N}}=1$, 2, and 3 supersymmetries. 
Since at large distances, the metrics look like
a direct product of three-dimensional Minkowski 
spacetime times an eight manifold, we motivate our
approach as a possible ``resolution'' 
of the eight-manifold singularity, by turning on the $F_4$ field.

We then studied the Penrose limits of the near horizon region 
in the metrics above, and arrived at 
a phenomenon that seems to be of general feature, 
namely, the pp-wave limit looks like the geometry 
that one would obtain by replacing the 
cone by a round sphere.  One can think of it as 
the limit is ``erasing'' the details 
of the particular manifolds that we consider. 
This gives rise to an interesting supersymmetry enhancement
phenomenon on the dual field theory which was already noted very recently in the
particular case of ${\cal N}=1$ super Yang-mills in four dimensions. 
Hence, we might conclude that likely there is 
a hidden ${\cal {N}}=8$ supersymmetry
which is present in the corresponding 
subsector of the dual SCFTs with less supersymmetries.

Finally, in an unrelated last section, we studied an RG flow between
two $AdS_6$ spaces. One of them preserves supersymmetry and it corresponds 
to a D4-D8 system. The second $AdS_6$ space
is non-supersymmetric. We have obtained numerically a kink solution interpolating
between the two vacua and commented on some gauge theory aspects 
like the dimensions of the operators that are inserted.

We would like to mention some open problems discussed in the paper. 
It would be interesting to find solutions, either in eight-dimensional 
supergravity or in M-theory, with an $F_4$ flux on the
supersymmetric four-cycle. This solution would make complete 
sense quantum mechanically, appart from being dual 
to a theory with Chern-Simons term. Another direction one would like
to explore is the more interesting case of four-dimensional gauge
theory arising from M-theory compactifications on $G_2$ holonomy manifolds.
In this case one would like to understand the dynamical singularity 
resolution mechanism analog to the one discussed in sections 4 and 5. 

~

{\bf Note added:}

While this paper was in preparation we received 
\cite{Loewy:2002hu} which overlaps with some results of section 4.
However their results have been obtained using a different approach.

~

\centerline{\bf Acknowledgements}

We would like to thank Ofer Aharony, Pascal Bain, Dan Freedman, Jerome Gauntlett, 
Gary Gibbons, Jaume Gomis, Joaquim Gomis, Amihay Hanany, 
Dario Martelli, Ventaka Nemani, Adam Ritz, James Sparks, David Tong
and Paul Townsend for helpful discussions. We would like to thank Angel Paredes 
and Alfonso Ramallo for valuable comments on the manuscript.
The work of U.G. and M.S. is supported in part by funds provided by the U.S. 
Department of Energy (D.O.E.) under cooperative research agreement 
$\#$DF-FC02-94ER40818. M.S. is also supported in part by Fundaci\'on 
Antorchas and The British Council. 
The work of C.N. is supported by PPARC. 

\newpage

{\large {\bf{Appendix A: $D=8$ Salam-Sezgin's gauged supergravity}}}

~

The bosonic part of the $D=11$ supergravity Lagrangian
\cite{Cremmer:1978km}
in tangent space is given by
\be
L_{D=11} = \frac{V}{4 \kappa^2} \, R_{(11)} - \frac{V}{48} F_{ABCD}
F^{ABCD}
+ \frac{2 \kappa}{144^2} \, \epsilon^{A_1, \cdot \cdot \cdot A_{11}} \,
F_{A_1, \cdot \cdot \cdot} \, F_{\cdot \cdot \cdot} \, V_{\cdot \cdot
\cdot A_{11}} \,\,\, 
\ee
We will borrow the notation of \cite{Salam:1984ft} where $V \equiv \det
V_M^A$
and the four-form field strength 
$F_{ABCD}=4 \, \partial_{[A} V_{BCD]} + 12 \, \omega_{[AB}^{\,\,\,\,
\,\,\,\,\,\, E} 
\, V_{CD]E}$. Indices $A, B=0, \cdot \cdot \cdot, 10$ are flat, while 
$M, N=0, \cdot \cdot \cdot, 10$ are curved. $V_{CDE}$ is the $D=11$
torsion-free
spin connection. The metric signature is taken to be mostly plus and we will
set $\kappa=1$ in 
what follows. 

By dimensional reduction on $S^3$, $ L_{D=11}$ becomes the $D=8$
Lagrangian obtained
by Salam and Sezgin \cite{Salam:1984ft}. The total field content of the theory
is
\ba
(g_{\mu\nu}, \, \psi^E_\mu, \, B_{\mu\nu\rho}, \, 
B_{\mu\nu\alpha}, \, A^\alpha_\mu, \, B_{\mu \alpha}, \, \chi^E_i, \,  
L_\alpha^i, \, B, \, \phi) \,\,\, , \nn
\ea
where $E=1, \, 2$ and $\alpha,\,\beta,\dots,=8,\, 9, \, 10$ and
$i,\,j,\dots=8,\, 9, \, 10$ 
are respectively curved and flat indices running over $S^3$, whereas 
 $\mu,\,\nu,\dots,=0,\dots,\,7$ and $a,\,b,\dots=0,\dots,\,7$ 
are curved and flat respectively.
$L_\alpha^i$ is a representative of the coset space $SL(3, \,R)/SO(3)$. 
Index $\alpha$ labels the global $SL(3, \,R)$ and the index $i$ labels the
composite $SO(3)$ \cite{Salam:1984ft}. Together with $B$ and $\phi$ these
constitute 7 scalars of the theory. 
One can further obtain a stable reduction down to the bosonic content,
\ba
(g_{\mu\nu}, \, B_{\mu\nu\rho}, \, A^\alpha_\mu, L_\alpha^i, \phi) 
\,\,\, , \nn
\ea
which is the background that we consider in our solutions. 
The bosonic part of $ L_{D=8}$ for this background is, 

\ba\label{8Dlagr}
e^{-1} \, L_{D=8} &=& \frac{1}{4} \, R_{(8)} - \frac{1}{4} \, e^{2 \phi}
\, F_{\mu\nu}^\alpha
F^{\mu\nu\beta} \, g_{\alpha \beta} - \frac{1}{4} \, P_{\mu ij} \, P^{\mu
ij} -
\frac{1}{2} \, \partial_\mu \phi \, \partial^\mu \phi \nn \\
&&- \frac{1}{16} \, g^2_c \, e^{-2 \phi} \,
(T_{ij} \, T^{ij} -\frac{1}{2} \, T^2) - \frac{1}{48} \, e^{2 \phi} \,
G_{\mu\nu\rho\sigma}
\, G^{\mu\nu\rho\sigma} \,\,\, .
\ea
Here there is a word of notation. 
We define 
$e \equiv \det e^\alpha_\mu$, $e^\alpha_\mu$ being the vielbein and
$R_{(8)}$ is the
curvature scalar in $D=8$. 

The $SU(2)$ two-form field strength is 
$F_{\mu\nu}^\alpha=\partial_\mu A^\alpha_\nu - \partial_\nu A^\alpha_\mu
+g_c \, \epsilon^{\alpha\beta\gamma} \, A^\beta_\mu A^\gamma_\nu$
and the kinetic term for $L^i_{\alpha}$ is the symmetric traceless part of
\be
P_{\mu i j} + Q_{\mu i j} = L_i^\alpha \, 
(\partial_\mu \, \delta_\alpha^\beta - g_c \, \epsilon^{\alpha\beta\gamma}
\,\, A^\gamma_\mu) 
\, L_{\beta j} \,\,\, ,
\ee
while $Q_{\mu i j}$ denotes the anti--symmetric part. 

We should mention that the dimensional reduction of the curvature scalar
implies the dimensional reduction of the spin connection. Since the latter is defined
in terms of the vielbein and its inverse, it is convenient to consider the reduction
ansatz for the $D=11$ vielbein to the $D=8$ 
\ba\label{uplift}
V_M^A \, = \, \left( \begin{array}{cc}
                     e^{- \phi/3} \, e^a_\mu & 0 \\
                   2 e^{2 \phi/3} \, A^\alpha_\mu \, L^i_\alpha & e^{2
\phi/3} \, L^i_\alpha
                                                   \end{array}
                                           \right) \,\,\, .
\ea
Using this gauge breaks the $SO(1,10)$ Lorentz group down to $SO(1,7) \times
SO(3)$.
Other quantities which appear in the scalar potential are 
$T^{ij} \equiv L^i_\alpha \, L^j_\beta \, \delta^{\alpha\beta}$ and 
$T\equiv\delta_{ij}T^{ij}$.  
Finally, we define the four-form field strength as
$G_{\mu\nu\rho\sigma}=(\partial_\mu \, B_{\nu\rho\sigma} + 3$
permutations$)$ 
$+ (2 \, F^\alpha_{\mu\nu} \, B_{\rho\sigma\alpha} + $ 5 permutations$)$.

The equations of motion for the system read, 
\ba
R_{\mu\nu} &=& P_\mu^{ij} \, P_\nu^{ij} + 2 \, \partial_\mu \phi \,
\partial_\nu \phi +
2 \, e^{2 \phi} \, F^i_{\mu\lambda} \, F^{\,\,\,\,\,\lambda, i}_\nu -
\frac{1}{3} \, g_{\mu\nu} \, \Box \phi \nn \\
\label{einstein}
&& + \frac{1}{3} \, e^{2 \phi} \,  
(G_{\mu\lambda\tau\sigma} \, G_\nu^{\,\,\,\,\,\lambda\tau\sigma} - 
\frac{1}{12} \, g_{\mu\nu} \, G_{\rho\lambda\tau\sigma} \,
G^{\rho\lambda\tau\sigma}) \,\,\, , \\
\label{einstein-eq}
\partial_\mu (\sqrt{-g} \, P^{\mu ij}) &=& - \frac{2}{3} \, \Box \phi \,
\delta^{ij} + 
e^{2 \phi} \, F^i_{\mu\nu} \, F^{\mu\nu, j} 
+ \frac{1}{36} \, e^{2 \phi} \, G_{\mu\nu\rho\lambda} \,
G^{\mu\nu\rho\lambda} \, \delta^{ij}
\nn \\
&&+ \frac{1}{2} \, g^2_c \, e^{-2 \phi} \, [T^i_n \, T^j_n - \frac{1}{2}
\, T \, T^{ij} 
- \frac{1}{2} \, \delta^{ij} (T_{mn} \, T^{mn} - \frac{1}{2} \, T^2)]
\,\,\, , \\
\label{eom-scalars}
\partial_\mu (\sqrt{-g} \, 
e^{2 \phi} \, F^{\mu\nu, i}) &=& - e^{2 \phi} \, P_\mu^{ij} \, F^{\mu\nu,
j}
- g_c \,\, g^{\mu\nu} \,\, \epsilon^{ijk} \, P^{jl}_\mu \, T_{kl} \,\,\, ,
\\
\label{non-abelianfield}
\partial_\mu (\sqrt{-g} \, e^{2 \phi} \, G^{\mu\nu\rho\sigma}) &=& 0
\,\,\, .
\label{G-field}
\ea

As shown in \cite{Salam:1984ft}, above equations of motion in $D=8$ 
can be uplifted to $D=11$ EOM derived
from the Cremmer-Julia-Scherk supergravity \cite{Cremmer:1978km}.
In particular, compactifications of $D=11$ supergravity containing either 
a three-dimensional manifold as a part of the seven-dimensional product
space, or as a
fiber space in a seven-dimensional fiber bundle, are also compactified
solutions of $D=8$ 
supergravity. One necessary condition in order to have these kind of
solutions
is that $SU(2)$ coupling constant $g_c$ is non-vanishing. To get 
supersymmetric solutions, the gauge connection of the normal bundle on a four-cycle $B^4$ 
(in the seven-dimensional $B^4 \times S^3$ space-time) 
must be equal to the spin connection of the spin bundle of $B^4$. 

~

{\large {\bf{Appendix B: General Solutions}}}

~

In this appendix we report about more general analytic solutions of BPS 
which contain (17) and (39) as special solutions. Some of these
solutions are \emph{physical} by Gubser's criterion as  we will demonstrate
below, hence they will correspond to RG flows in the dual gauge theory.
However, field theory interpretation is not clear to us at the
moment of writing this manuscript.

For the case of the ${\cal {N}}=1$ system, as a first step to obtain
the general solution, let us define the field $x\equiv
20 \, e^{2\phi-2h}$. The differential equation for $x$ in the
variable 
$d/dt= e^{\phi}d/dr$, is solved as, 
\be\label{xsol}
x(t)=\frac{1}{1+a e^{-t/2}}.
\ee  
where $a$ is an integration constant. Note that $a=0$ corresponds to the
special solution (17). $a<0$ case will be unacceptable by Gubser's
criterion, hence we consider $a>0$ for which, from
 (\ref{fprime})-(\ref{hprime}), we can find the general solution, 
\ba
h(t) &=& \frac{1}{4}\log\bigg(\frac{4\Lm}{5}(c-I(v))\bigg)+\frac{t}{8}
+\frac{1}{5}\log(a+e^{t/2})-\frac{9}{20}\log(a),\\
\phi(t)&=& h(t)-\frac{1}{2}\log\bigg(20(1+ae^{-t/2})\bigg).
\ea
where $I(t)$ is defined as,
\ba
I(v) &=& \Bigg\{\frac{5v}{1-v^5}-
4\log(1-v)\nn\\
{}& & +2\sqrt{2(5+\sqrt{5})}\arctan\left(\frac{4v-(\sqrt{5}-1)}
{\sqrt{2(5+\sqrt{5})}}\right)\nn\\
{}& & + 2\sqrt{2(5-\sqrt{5})}\arctan\left(\frac{4v+(\sqrt{5}+1)}
{\sqrt{2(5-\sqrt{5})}}\right)\nn\\
{}& & -(\sqrt{5}-1)\log(v^2-\frac{1}{2}(\sqrt{5}-1)v+1)\nn\\
{}& & +(\sqrt{5}+1)\log(v^2+\frac{1}{2}(\sqrt{5}+1)v+1)\Bigg\} \,\,\, ,
\ea
where $v=(\frac{e^{t/2}}{a}+1)^{1/5}$.

In order to obtain the fixed point solution (\ref{metricfp}) 
in the limit $t\to\infty$,
the integration constant of $h(t)$ should be chosen as
$c=\pi(\sqrt{2(5-\sqrt{5})}+\sqrt{2(5+\sqrt{5})})$.

On the other extremun of the flow, $t\to -\infty$ we can read the 
asymptotic expansions of the fields and the eleven-dimensional solutions
to be
\beq
h\approx \frac{1}{4} \, \log(4 \Lambda),\;\;\; \phi \approx t/4 + 
\frac{1}{4} \, \log(\Lambda/100),\;\; f\approx t/4 \,\,\, ,
\eeq
and in the variable $d\rho= \frac{\Lambda}{100} \, e^{t/6} \, dt$ such that $t\to
-\infty$ is $\rho=0$
\beq
ds_{11}^2 =\rho^2 \, dx_{1,2}^2 + \frac{6}{\rho} \, \left(\frac{80^2
\Lambda^3}{100}\right)^{1/6} \, d\Omega_{4}^2 +
 \frac{\rho^2}{36} \, (\omega^i-A^i)^2 + d\rho^2 \,\,\, ,
\label{metricappendix}
\eeq
and, in flat indexes, 
\beq
F_{xyt\rho}\approx 1/\rho \,\,\, .
\eeq
A fact that we would like to remark is that, if we do not impose
the integration constant to be
$c=\pi(\sqrt{2(5-\sqrt{5})}+\sqrt{2(5+\sqrt{5})})$
the solution ``misses'' the fixed point and  
for large values of the radial variable $t$ it asymptotically approaches
a cone over weak $G_2$ manifold, having an expression as
Eq.(\ref{spin7cone}).

The solution above has a curvature singularity for small values of the
radial coordinate $\rho$.  We should be able to decide whether we should
accept 
or not the singular behaviour. In order to do that, we can ``integrate
out'' the degrees of freedom
living on the four-sphere and the gauge fields in the eight-dimensional 
supergravity Lagrangian and transform the system into one of gravity plus
scalars.

Having done this, we find a four-dimensional Lagrangian that reads
\beq
L=\sqrt{g_4} \, [ R + T  - V_{eff}] \,\,\, ,
\label{lageff}
\eeq
where, $T$ denotes kinetic terms for the fields $\phi,h$ and $V_{eff}$ is
given by
\beq
V_{eff}= -3 e^{-6h} -\frac{\Lambda^2}{2}e^{-12h-2\phi} 
- \frac{3}{32}e^{-2\phi-4 h}+ \frac{9}{4}e^{2\phi-8h} \,\,\, .
\eeq
We can see that near 
this singularity, this potential is bounded above, this renders the
singularity to be \emph{acceptable} according to the criterion introduced in
\cite{Gubser:2000nd}. We conclude that UV of the dual field theory is
at $t=+\infty$ for which the geometry is $AdS_4\times\tilde{S_7}$. IR
is at $t=-\infty$ for which one obtains above asymptotic geometry. 
Note that in $\rho\,\to\,0$ limit of (\ref{metricappendix}), 
$S^4$ radius becomes much larger than size of the domain--wall. 
This fact renders field theory interpretation difficult in
eleven dimensions. However, considering the eight-dimensional 
supergravity problem, one does not run
into trouble with the nature of the singularity at $t \to -\infty$ which
is acceptable by above criteria and corresponds to the IR limit of the
corresponding field theory. Therefore, one can conclude that
$t\to\,\infty$ limit ($AdS_4\times S_4$) corresponds to UV
and $t\to\,-\infty$ limit of eight-dimensional super-kink corresponds to IR of the dual 
${\cal {N}}=1$ SYM in three dimensions.         

For the case of ${\cal {N}}=2$, similarly we first introduce the
field, 
$x\equiv 4e^{2\phi-2h+2\lm}$ which can be solved in the
variable $d/dt= e^{\phi+4\lm}d/dr$ as in (\ref{xsol}). 
A more general solution than (39) can be obtained however by keeping 
$x=1$ frozen, but turning on $\lm$ field: 

\ba
\lm(t) &=& \frac{1}{6}\log\left(\frac{2 e^{2t}}{e^{2t}-1}\right) \,\,\,
, \nn \\
e^{4 h(t) - 4\lambda(t)} &=& \Lm\left(
\log\left(\frac{e^{t/2}+1}{e^{t/2}-1}\right)
+2\arctan(e^{t/2})-C\right)e^{-t/2}(e^{2t}-1)
\,\,\, \nn \\
\phi(t)&=&h(t)-\lm(t)-\log(2) \,\,\, \nn \\
f(t) &=& -\frac{t}{24}+\frac{1}{12}\log\left(e^{2t}-1\right)-
\frac{1}{4}\log\left(
\log\left(\frac{e^{t/2}+1}{e^{t/2}-1}\right)
+2\arctan(e^{t/2})-C\right)\nn   
\ea
where for convinience we fixed the integration constant for $\lambda$.
If constant $C$ above is chosen as $C=\pi$, then the solution 
reaches the fixed point solution of Eq.(\ref{q111}).

On the other hand, if we do not impose this value for the integration
constant, again as in the ${\cal {N}}=1$ case, the solution misses the
fixed point.
Then, for large values of the radial variable $t$ 
the functions $f, \,\, h, \,\, \lambda, \,\, \phi$ approach
\beq
\lambda=1/6 \, \log(2),\;\; \phi\approx h\approx 3f\approx 3/8t \,\,\, .
\eeq
Thus leading to a metric of the form $E^{2,1} \times CY4$ like in
Eq.(\ref{metricsing}).

On the other extremun of the flow ($t\to 0$), 
there is a curvature singularity as in ${\cal {N}}=1$ case.  
We can also define a similar effective Lagrangian, but in this case the
kinetic terms
will include the field $\lambda$. The effective potential reads
\beq
V_{eff}= - e^{-6 h} -\frac{\Lambda^2}{2}e^{-2\phi-12 h} +
e^{2\phi-8h-4\lambda} + \frac{g^2}{32}
e^{-2\phi-4h}(e^{-8\lambda} - 2 e^{-2\lambda}) \,\,\, .
\eeq
If we analyze the behaviour of this effective potential 
near the IR, we see that it is again bounded above, rendering
the singularity \emph{acceptable}. One can conclude that
$t\to\,\infty$ limit ($AdS_4\times S_2\times S_2$) corresponds to UV
and $t\to\,0$ limit of eight-dimensional super-kink corresponds to IR of the dual 
${\cal {N}}=2$ SYM in three dimensions.    
  
Yet a most general solution  can be found analytically 
by turning on $x$ defined above 
(solution will be Eq.(\ref{xsol})). It is also acceptable 
by means Gubser's criterion for some ranges of the
integration constants.  

~

{\large {\bf{Appendix C: Orbits of BPS equations: ${\cal {N}}=1$ case}}}

~

Although we were able to find analytical solutions of the
BPS equations for the situations studied in this paper, it is 
interesting to see how this solutions behave as orbits
of the BPS equations. Here we show the orbits for the 
case preserving two supercharges, {\it i.e.} our first example.
The other cases for ${\cal {N}}=2$ and ${\cal {N}}=3$ are 
similar since for such cases we have considered situations with only
irrelevant deformations, such that for instance $\lambda$ is fixed so 
that at the end we can reduce the analysis to one orbit equation as Eq.(\ref{ORBITS}).

For the present case we consider the orbits of Eq.(\ref{ORBITS}).
We recall that we are doing a similar analysis as in \cite{Acharya:2000mu}.
In figure 6 we plot some numerical solutions of Eq.(\ref{ORBITS}).
The straight line $H=20 s$ represents a collection of IR fixed points
if we consider flowing down from a Spin(7) manifold at UV, for different values
of $\Lambda$. As an example
we plot a short solid line segment indicating this kind of flow
that is the same as described in figure 2 in the body of the paper. The rest of
solid lines correspond to solutions for non-vanishing $\Lambda$, which are ``missing''
IR fixed points.
At UV all of these curves approach $H= 20 s$ leading to the metric
of $E^{2,1} \times Spin(7)$. All them approach the origin with the 
same slope. They correspond to some of the solutions included in 
Appendix B. Indeed, the almost vertical solid line around the
point labeled as IR f.p. is one of the solutions introduced in Appendix B 
for which we do not
have a clear physical interpretation at the present. Particularly,
for this solution the IR/UV limits are the opposite as labeled
in figure 6.

The dashed line represents a pure-metric solution, {\it i.e.} when
$\Lambda=0$.

\vspace{0.5cm}
\begin{center}
\epsfig{file=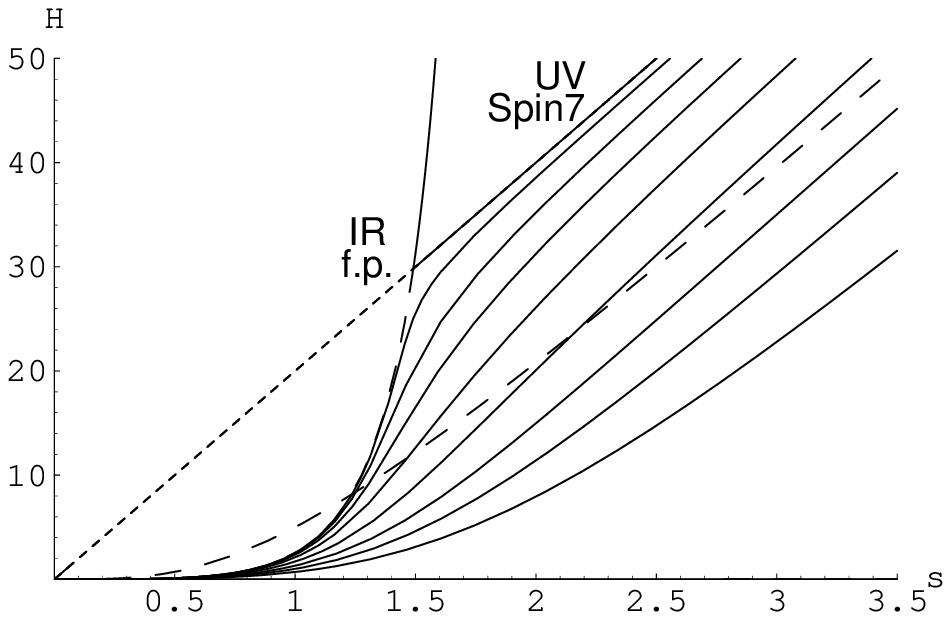, width=9cm}
\end{center}
\vspace{0.5cm}
\baselineskip=13pt
\centerline{\small{Figure 6: Orbits for BPS equations for ${\cal {N}}=1$ case.}}

\vspace{0.5cm}

\baselineskip=15.5pt

~

{\large {\bf{Appendix D: Notation for the left-invariant one-forms}}}

~

In the definition of the squashed seven-sphere we have used the
left-invariant one-forms as follows. For the three-sphere in $S^4$
we used
\bea
\sigma^1 & = & \cos\psi \, d\theta + \sin\psi \, \sin\theta \, d\phi \,\,\, , \nn \\
\sigma^2 & = & \sin\psi \, d\theta - \cos\psi \, \sin\theta \, d\phi \,\,\, , \nn \\
\sigma^3 & = & d\psi            +             \cos\theta \, d\phi \,\,\, ,  
\eea
while for the $SU(2)$ group manifold we have used
\bea
\omega^1 & = & \cos\gamma \, d\alpha + \sin\gamma \, \sin\alpha \, d\beta \,\,\, , \nn \\
\omega^2 & = & \sin\gamma \, d\alpha - \cos\gamma \, \sin\alpha \, d\beta \,\,\, , \nn \\ 
\omega^3 & = &    d\gamma            +               \cos\alpha \, d\beta \,\,\, .
\eea

\end{document}